\documentclass{article}

\usepackage{microtype}
\usepackage{graphicx}
\usepackage{subcaption}
\usepackage{adjustbox}
\usepackage{booktabs} 
\usepackage{multirow} 
\usepackage{array}

\usepackage{hyperref}

\usepackage{enumitem} 
\usepackage{gensymb} 
\usepackage[version=4]{mhchem} 
\usepackage{siunitx} 
\usepackage{booktabs}

\usepackage[preprint]{icml2026}

\usepackage{amsmath}
\usepackage{amssymb}
\usepackage{mathtools}
\usepackage{amsthm}

\usepackage[capitalize,noabbrev]{cleveref}

\theoremstyle{plain}

\theoremstyle{definition}

\theoremstyle{remark}

\usepackage[textsize=tiny]{todonotes}

\icmltitlerunning{LitXBench: A Benchmark for Extracting Experiments from Scientific Literature}

\usepackage{listings}
\usepackage{xcolor}

\usepackage[most]{tcolorbox}
\usepackage{listings}
\usepackage{lipsum}

\tcbuselibrary{skins,breakable,listings}

\newtcblisting{prompt}[1][]{
    enhanced,
    breakable,
    colback=white,
    colframe=gray!80!black,
    boxrule=1.2pt, 
    arc=1.5mm,
    borderline={1.2pt}{0pt}{gray!80!black},
    skin first=enhanced,
    skin middle=enhanced,
    skin last=enhanced,
    pad at break=3mm,
    top=3mm,
    bottom=3mm,
    left=3mm,
    right=3mm,
    listing only, 
    listing engine=listings,
    listing options={
        basicstyle=\ttfamily\small,
        breaklines=true,
        columns=flexible,
        keepspaces=true,
        frame=none, 
    },
    #1
}

\usepackage[framemethod=TikZ]{mdframed} 
\usepackage{caption}

\definecolor{codebackground}{RGB}{245, 245, 247}
\definecolor{keywordcolor}{RGB}{0, 51, 153}
\definecolor{stringcolor}{RGB}{153, 0, 51}
\definecolor{commentcolor}{RGB}{102, 102, 102}

\begin{document}

\twocolumn[
  \icmltitle{LitXBench: A Benchmark for Extracting Experiments from Scientific Literature}

\begin{center}
\begin{tabular}{cc}
{\large Curtis Chong} & {\large Jorge Colindres} \\
{\normalsize Radical AI} & {\normalsize Radical AI} \\
{\ttfamily cchong@radical-ai.com} & {\ttfamily jc@radical-ai.com}
\end{tabular}
\end{center}

\vskip 0.23in
]

\begin{abstract}

  Aggregating experimental data from papers enables materials scientists to build better property prediction models and to facilitate scientific discovery. Recently, interest has grown in extracting not only single material properties but also entire experimental measurements. To support this shift, we introduce LitXBench, a framework for benchmarking methods that extract experiments from literature. We also present LitXAlloy, a dense benchmark comprising 1426 total measurements from 19 alloy papers. By storing the benchmark's entries as Python objects, rather than CSV or JSON, we improve the auditability of the data and enable programmatic validation. We find that frontier language models, such as Gemini 3.1 Pro Preview, outperform existing multi-turn extraction pipelines by up to 0.37 F1. Our results suggest that this performance gap arises because extraction pipelines associate measurements with compositions rather than the processing steps that define a material.
  
\end{abstract}

\begin{figure}[t]
    \centering
    \includegraphics[width=\columnwidth]{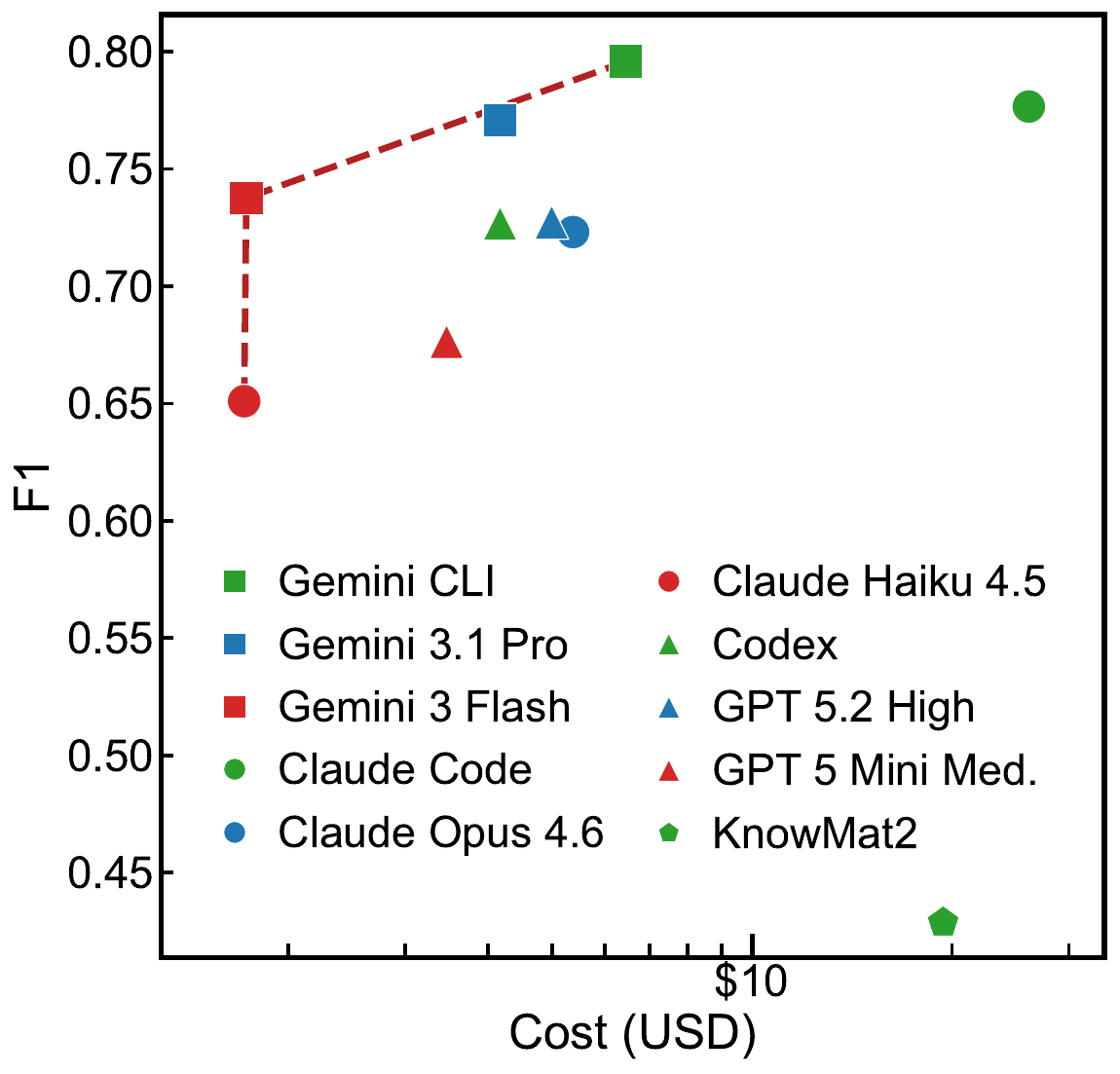}
    \caption{Pareto front of experiment extraction methods. The cost is the total cost for all papers in LitXAlloy.}
    \label{fig:myimage}
\end{figure}

\section{Introduction}

Experimental materials science datasets anchor computational materials science to real-world validation. They are relevant to validate machine learning pipelines \cite{dunn2020benchmarking}, train property prediction models \cite{ward2016general, chen2021leveraging}, and to collect comprehensive data across an entire materials space, for example, to construct phase diagrams \cite{saunders1998calphad}. Efforts such as Optimade \cite{andersen2021optimade} and the Materials Project \cite{Jain2013} address the scarcity of experimental data by providing a standardized interface for accessing material properties. There are also websites designed for scientists to record their experiments in databases \cite{nims_matnavi, matweb}. However, community-aggregated databases have limited expressiveness, as they focus on basic material properties and lack detailed experimental data necessary for simulations \cite{rumor2022need}.

\begin{figure*}[h]
\centering
\includegraphics[width=0.8\textwidth]{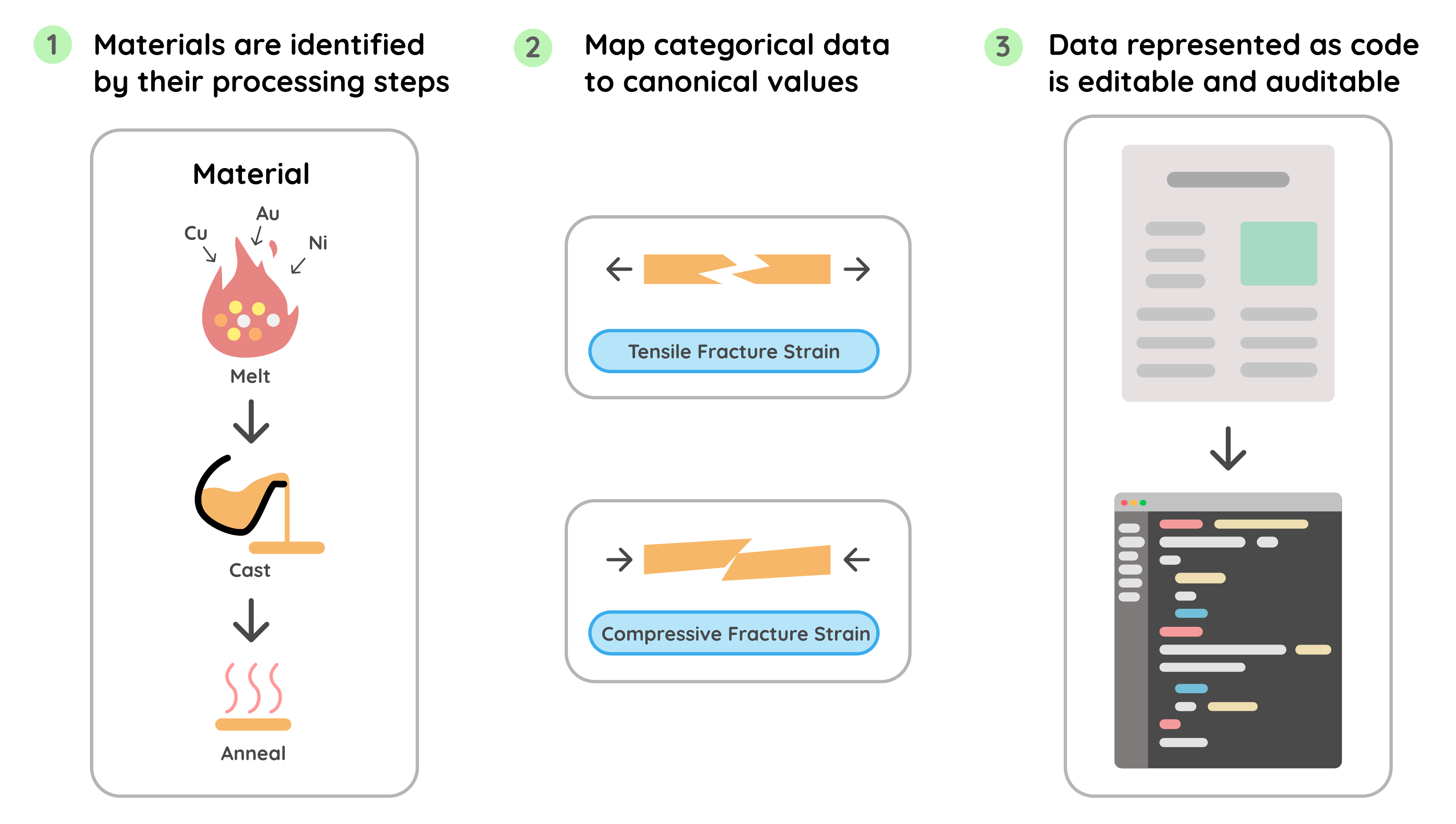}
\caption{\textbf{LitXBench Principles for Accurate Extraction and Benchmarking.} (1) To accurately capture a material's properties, measurements must be linked to its processing lineage, rather than just its composition. (2) Categorical values should be mapped to canonical identifiers to disambiguate similar values, as multiple papers may reference different properties with the same term. (3) Extracted materials are more editable and auditable when represented as code, reducing errors in the benchmark.}
\label{fig:takeaways}
\end{figure*}

Therefore, a more practical approach is to mine experiments from literature, as researchers can control the amount and fidelity of data acquired. Although manually aggregated datasets exist, they are impractical to scale, as for example, the OBELiX \cite{therrien2026obelix} and MPEA \cite{borg2020expanded} datasets contain only $\sim$600 and 1545 entries, respectively. These datasets are particularly small when subdividing for materials with specific processing conditions or properties. For example, in the MPEA dataset, there are only 95 entries made from powders, and out of those, only 10 have an observed BCC phase. Thus, there is great interest in developing automated data-mining workflows to collect larger datasets from the literature \cite{ward2018matminer, hong2021challenges, mahbub2020text}.

Rule-based methods are at the heart of early scientific information extraction pipelines such as ChemDataExtractor \cite{swain2016chemdataextractor} and LeadMine \cite{lowe2016efficient}. However, they often make mistakes, because heuristics are too brittle to connect information dispersed across papers \cite{atagong2025review, kim2019distilling}. Information is also inconsistently described between papers, as different aliases may be used for the same concept \cite{cho2017method}. As seen in one paper, `` `Na0.8' is used as the abbreviation for `P2 - Na$_{0.8}$Ni$_{0.1}$Mn$_{0.8}$Fe$_{0.1}$O$_2$' leading to incomplete identification of chemical entities'' \cite{gou2024document}. This example shows that traditional experiment extraction methods lack the intelligence to bind disparate concepts. Such contextual understanding is also important for identifying misnomers and typos \cite{flor2012using}. Since humans take hours to comprehend each work, fully understanding a paper requires deep reasoning \cite{kim2019distilling}, and hence heuristical rules for extraction are insufficient.

With the recent introduction of Large Language Models (LLMs), data mining methods can now integrate deeper understanding, shifting the field away from Named Entity Recognition \cite{jensen2019machine, cruse2022text, he2020similarity, zhang2023literature} to extracting synthesized materials and their properties \cite{dagdelen2024structured, liu2025thermodynamic, polak2024extracting, fleyhag_scase, khalighinejad2024extracting, li2025exploring, sayeed2025knowmat, wang2025reliable}. These works currently rely on their own benchmarks, which lack exact ground-truth annotations \cite{lederbauer2025lemat}, do not have a publicly released eval-set \cite{sayeed2025knowmat}, or do not go into sufficient detail for their manual annotation \cite{liu2025thermodynamic, khalighinejad2024extracting}. As discussed later in \autoref{appendix:benchmark_quality}, creating a trustworthy validation set is challenging. Thus, without sufficient justification or auditability of the validation sets, human-extracted values are, by themselves, untrustworthy. The lack of accurate ground-truth datasets precludes a standardized and accurate benchmark for extracting experimental material data, making comparisons between works challenging. To make extraction methods comparable, we introduce LitXBench, a framework to benchmark experimental data extraction methods.

Our contributions are fourfold. First, we introduce LitXAlloy, a high-quality LitXBench-based benchmark derived from 19 alloy papers, with values hand-extracted and extensively reviewed by both humans and LLMs. Second, we benchmark frontier LLMs on LitXAlloy and show that extraction accuracy substantially improves when models are tasked to extract properties for each material (which is identified by its synthesis process) rather than for each composition. Third, we show that categorical values are best distinguished when mapped to unique canonical identifiers. Fourth, we present an editable and auditable code-based data model to represent experiments.

\section{Methodology}
\subsection*{Experiment Extraction Problem Description}

Formally, a paper contains a set of synthesized materials $M$. Each material $m_i$ is created from a synthesis process $p_i$ and has measurements $x_i$. These measurements are the material's characterization and property measurements. The experiment extraction task is to output all measurements $x_i$ for each material $m_i$. Given the one-to-one relationship between each of these values, the result of experiment extraction can be represented as a list of tuples $(m_i, p_i, x_i)$.

Note that materials $m_i$ are not compositions. Since compositions are measured values, there can be multiple composition measurements for each material. For example, the composition measured by a scale, energy-dispersive X-ray spectroscopy, or optical emission spectrometry is recorded in $x_i$ for each material $m_i$. Moreover, identifying materials by composition can incorrectly map two materials to the same object. To illustrate this problem, consider \cite{zhang2019gradient}, where two materials were made with the same composition of $\mathrm{Fe}_{24.1}\mathrm{Co}_{24.1}\mathrm{Cr}_{24.1}\mathrm{Ni}_{24.1}\mathrm{Mo}_{3.6}$ but with different pressures during spark plasma sintering. Failure to disambiguate the two materials creates an invalid one-to-many mapping between the composition and the properties. Thus, the extraction must identify samples with unique processing conditions as distinct materials.

\subsection*{Benchmark Constituents}

LitXBench is designed to evaluate models on the experiment extraction problem. It is compiled from 19 papers. Eighteen of these \cite{yang2012microstructure, gludovatz2016exceptional, shi2019enhanced, rickman2019materials, nene2017enhanced, sanchez2019design, el2019effect, xu2018microstructure, chen2014microstructures, yao2016monbtav, tseng2018effects, zyka2019microstructure, sun2019phases, haas2019microstructure, tan2019effect, zhang2019effects, jia2019novel, zhang2019gradient} are open-access papers from the MPEA dataset \cite{borg2020expanded}; with an additional open-access paper on Ni-based superalloys \cite{kanetas2020ebsd}. The Ni-based superalloy paper was selected because of its complex synthesis process and unique experimental measurements. Although 19 appears to be a small number, it is comprehensive considering that LitXAlloy contains 1426 datapoints. Furthermore, increasing the dataset would divide annotator attention, reducing the benchmark's accuracy. Further details of LitXAlloy, including its statistics, construction quality, and comparison with the MPEA dataset, are in \autoref{appendix:benchmark_details}.

\subsection*{Canonical Values for Ontological Disambiguation} \label{sec:canonical_val_criteria}
In LitXBench, all extracted categorical properties are mapped to a global identifier that identifies each property. These identifiers are referred to as canonical values. The main benefit of canonical values is to prevent alias collisions. For example, the fracture strain in \cite{yao2016monbtav} is reported from a compression test, whereas in \cite{zhang2019effects} it is reported from a tensile test. Without canonical values to disambiguate these properties, an LLM may map these incomparable properties to the same ``fracture strain" string. Canonical values can also normalize terms that differ due to errors or by differing conventions. Lastly, canonicalization is not limited to material properties; all categorical fields, ranging from process conditions to microstructure types, require it to resolve ambiguity.

\begin{figure*}[h]
\centering
\vspace{2pt}
\includegraphics[width=1\textwidth]{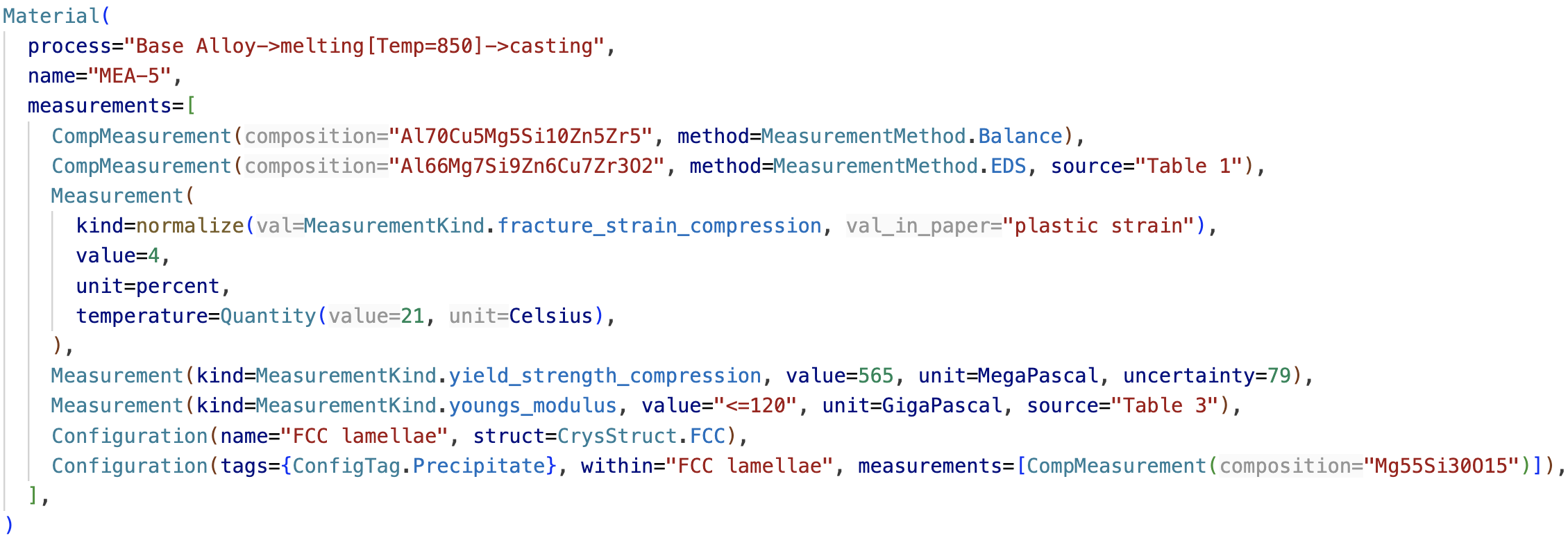}
\caption{Schema of extracted materials in LitXAlloy. Each material is identified by its process steps, which are outlined by the arrow notation. Measurements performed on the material follow. CompMeasurements are various composition measurements performed on the sample. Configuration measurements correlate to microstructure and other features typically visible through an electron microscope. Further schema specification details are in \autoref{appendix:data_models}.}
\label{fig:sample_code_example}
\end{figure*}

\subsection*{Code-Based Benchmark Representation}

Human-extracted datasets, such as ChemX \cite{vepreva2025benchmarking}, Zhang \cite{zhang2010diffusion}, and MPEA \cite{borg2020expanded}, are often used as benchmarks for extraction \cite{ghosh2024toward}. However, these works are infrequently updated. As demonstrated by \cite{northcutt2021pervasive}, many widely used benchmarks contain substantial errors. Correcting such errors is important because \cite{jin2026pervasive} demonstrates that corrections can shift results such as leaderboard rankings by as many as $\pm9$ positions. Despite this, datasets are rarely updated after release, as \cite{rondina2023completeness} finds that only 30 of 100 datasets were recently updated, and none of them have an erratum. A reliable benchmark should therefore exhibit two properties: it should be easy to edit when mistakes are discovered, and it should be auditable so future annotators can understand why a value was recorded. LitXBench satisfies these requirements by representing materials as code.

Representing the benchmark as code makes it easier to correct errors. By storing information as the Python objects shown in \autoref{fig:sample_code_example}, LitXBench inherits the robust type checking and syntactic guarantees of modern IDEs. Edited values maintain high human-readability, as external tools (such as spreadsheet viewers) are not needed. Go-to-definition commands can also be used to quickly locate enum definitions, which define the ontology's canonical values. Unlike one-time CSV data releases, benchmarks released as code inherit the benefits of open-source software, inviting contributors to quickly amend errors. Overall, LitXBench reflects a broader trend in computational materials science exemplified by packages such as Atomate2 \cite{ganose2025atomate2}, where code is used to facilitate collaboration.

Code also provides provenance and invites debate. Papers often omit critical details, leaving readers to infer missing information. As a result, extracted measurements are contested, even among experts. Since most data-mined datasets lack source citations for extracted values, it is challenging for external moderators to verify correctness or understand ambiguous cases. As demonstrated in the field study conducted in \cite{kuo2024wikibench}, when datasets are updated by the broader community, they "effectively capture community consensus, disagreement, and collective uncertainty". LitXBench enables similar levels of collaboration by supplying a \texttt{source} field on measurements. By indicating where a value originated, annotators can justify assumptions made during extraction with in-depth explanations. This allows LitXBench users to assess the accuracy of each entry and improve their extraction method accordingly. Additionally, annotators can comment out extracted measurements until there is sufficient consensus to uncomment. These comments show that the value was extracted with care, but that the annotators are uncertain. More broadly, commented-out measurements can also inform maintainers that the current data model does not adequately support them. Above all, comments can explain why values were omitted, preventing future contributors from incorrectly overriding seemingly anomalous values or adding incorrect measurements.

\begin{figure}[t]
\centering
\vspace{2pt}
\includegraphics[width=0.5\textwidth]{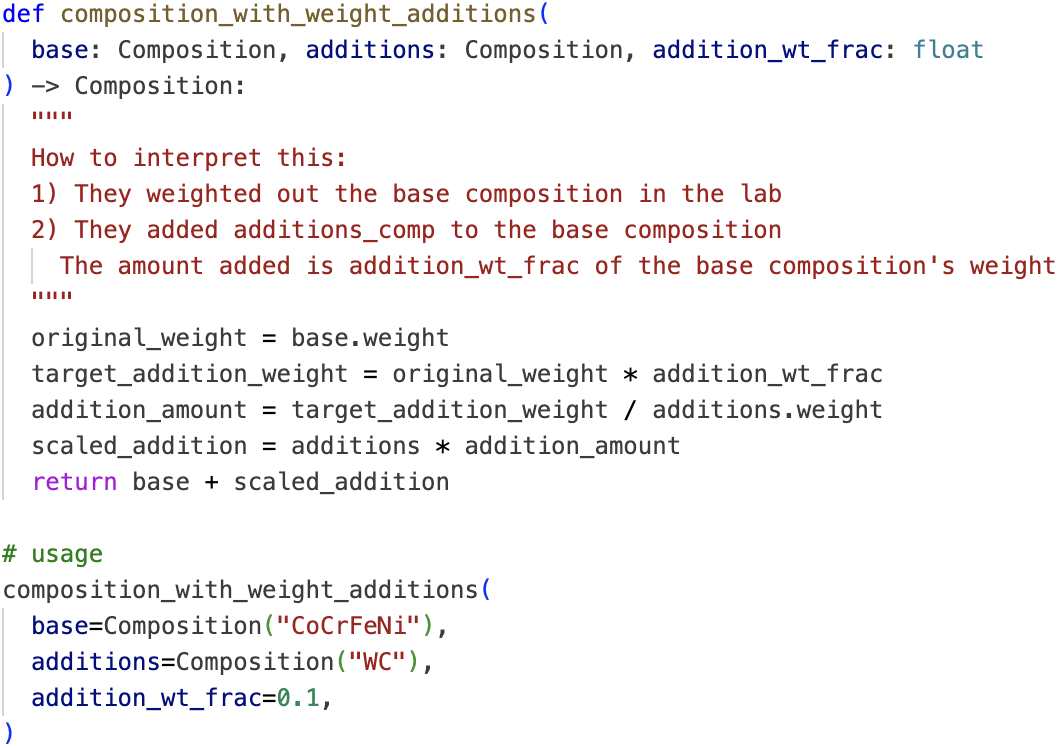}
\caption{Helper function which calculates a composition in which the weights of the additions are based on the weight of the base alloy. Since the code is self-documenting, this preserves the interpretability of the composition reported in the paper when compared to its nominal composition of $\mathrm{CoCrFeNiW}_{0.12}\mathrm{C}_{0.12}$}
\label{fig:comp_wt_additions_helper}
\end{figure}

Code also enables helper functions to make normalization auditable. For example, in \cite{xu2018microstructure}, equiatomic Tungsten Carbide particles were added that constitute 10\% by weight of the base alloy. Converting that composition to atomic percent requires many steps, so the helper function in \autoref{fig:comp_wt_additions_helper} is used to normalize the composition. Consequently, auditing the composition shifts the burden from recalculating the nominal composition by hand to understanding the helper function’s implementation. In contrast, values in the MPEA dataset are stored in a CSV, requiring compositions to be normalized in advance, which obscures their reported form in the paper, leading to uncaught errors.

\begin{table*}[ht]
\centering
\small
\begin{tabular}{l ccc cccc cc}
\toprule
& \multicolumn{3}{c}{\textbf{Overall}} & \multicolumn{4}{c}{\textbf{Per-Category F1 Scores}} & \multicolumn{2}{c}{\textbf{Efficiency}} \\
\cmidrule(lr){2-4} \cmidrule(lr){5-8} \cmidrule(lr){9-10}
\textbf{Method} & Prec. & Rec. & F1 & Meas. & Proc. & Mat. & Config & Attempts & Total Cost (USD)\\
\midrule
KnowMat2 (GPT-5.2 High) & 0.52 & 0.43 & $0.43 \pm 0.06$ & 0.27 & 0.67 & 0.65 & 0.17 & - & 19.40 \\
\cmidrule(lr){1-10}
Claude Haiku 4.5 & 0.64 & 0.69 & $0.65 \pm 0.01$ & 0.51 & 0.84 & 0.94 & 0.38 & 2.21 & \textbf{1.72} \\
GPT-5 Mini Medium & 0.67 & 0.70 & $0.68 \pm 0.04$ & 0.52 & 0.84 & 0.94 & 0.41 & 2.49 & 3.47 \\
Gemini 3 Flash Preview & 0.74 & 0.76 & $0.74 \pm 0.05$ & 0.61 & 0.86 & 0.97 & 0.52 & 2.58 & 1.73 \\
\cmidrule(lr){1-10}
Claude Opus 4.6 & 0.75 & 0.73 & $0.72 \pm 0.04$ & 0.62 & 0.86 & 0.91 & 0.54 & 1.53 & 5.37 \\
GPT-5.2 High & 0.71 & 0.77 & $0.73 \pm 0.02$ & 0.65 & 0.85 & 0.97 & 0.49 & 1.46 & 4.99 \\
Gemini 3.1 Pro Preview & 0.79 & 0.77 & $0.77 \pm 0.03$ & 0.71 & 0.83 & 0.96 & 0.60 & 1.51 & 4.17 \\
\cmidrule(lr){1-10}
Claude Code (Opus 4.6) & \textbf{0.81} & 0.77 & $0.78 \pm 0.01$ & 0.71 & \textbf{0.88} & 0.94 & 0.56 & 1.26 & 26.11 \\
Codex (GPT-5.2 Codex High)& 0.76 & 0.72 & $0.73 \pm 0.01$ & 0.67 & 0.82 & 0.95 & 0.52 & 1.49 & 4.17 \\
Gemini CLI (3.1 Pro Preview) & 0.80 & \textbf{0.81} & $\mathbf{0.80 \pm 0.04}$ & \textbf{0.74} & 0.84 & \textbf{0.98} & \textbf{0.68} & 2.47 & 6.46 \\
\bottomrule
\end{tabular}
\vspace{6pt}
\caption{Model performance on the experiment extraction task. The per-category F1 scores correspond to model performance on extracting measurement values (Meas.), process conditions (Proc.), the set of materials (Mat.), and the set of microstructure (Config). The score weight contributions are: Meas.=0.5, Proc.=0.2, Mat.=0.15, Config=0.15.}
\label{tab:zero_shot_code}
\end{table*}

Unlike data stored in CSV or JSON formats, code natively provides compile-time and run-time validation guarantees, reducing human error when amendments are made to the benchmark. As extraction pipelines increase in complexity from single LLM calls to multi-turn agentic workflows \cite{schilling2025text}, validation becomes important because they teach the LLM to retry and correct its mistakes. These validation errors guide the LLM by surgically identifying the offending line. Our experiments in \autoref{appendix:json_extraction} find that exceptions referencing offending objects reduce the number of attempts to produce a valid extraction schema. Beyond syntax validation, LitXBench enforces consistency checks, such as ensuring that there are no name collisions or graph cycles (a material cannot depend on itself as a precursor). Furthermore, code facilitates semantic validation. For example, in LitXAlloy, all melting processing events must be followed by a casting event. Compositions are also validated to ensure that they sum up to 100\%. These are more than compile-time checks; they are alloy-specific checks that must be satisfied for the extraction to be valid. Although these checks could be implemented in JSON-like extraction schemas, the output would still need to be converted into code objects for validation. Since correcting exceptions in code is a common task in LLM post-training \cite{liu2023rltf}, exceptions are the natural medium to indicate validation issues.

\section*{Experimental Setup}\label{section:experiment_setup}

Each experiment was performed on transcribed text from Mistral OCR 3 \cite{mistral_ocr3_2025}. The figures from the transcription are excluded from LLM prompts, as extracting information from images is beyond the scope of this study.\footnote{Manually extracting information from figures is challenging because charts require external tools such as WebPlotDigitizer \cite{WebPlotDigitizer} to properly identify precise numerical values. For example, bar charts requires users to align the top of the bar with an axis tick to calculate the bar's value.} This is an acceptable compromise, as most information is reiterated in the paper's text. Furthermore, authors often highlight important findings from figures in the text, thereby minimizing information loss.
Tables are expressed as text in markdown and are visible to the models.

The models are tasked with extracting materials using the schema specified in \autoref{appendix:data_models}. Since the output format is code, we evaluate the models via API access\footnote{Models are accessed using Pydantic AI \cite{PydanticAI2024}} and their equivalent coding CLIs. Uncertainty scores are calculated by taking the 95\% confidence interval of the Student's t-distribution of three runs. Prompts for the experiment extraction task are under \autoref{appendix:experiment_extraction_prompts}, and KnowMat2 \cite{sayeed2025knowmat} evaluation details are in \autoref{appendix:knowmat2}. Similar to existing work \cite{khalighinejad2024extracting}, the Hungarian algorithm \cite{kuhn1955hungarian} is first used to find the maximum bipartite match between the set of extracted materials $M_\text{extracted}$ and target materials $M_\text{target}$. Once the optimal assignments are determined for both sets of materials, the F1 score is used to evaluate extraction accuracy, as in prior work on scientific extraction \cite{li2025exploring}. The cost functions used to calculate the Hungarian score and overall F1 score are explained in \autoref{appendix:scoring_cost}.

\section{Results} \label{section:results}

\autoref{tab:zero_shot_code} shows the performance of frontier LLMs and the KnowMat2 extraction workflow on experiment extraction. While there are significant performance gains between the smaller and larger models, the Gemini series deviates from this pattern with consistently high scores. Across the board, agentic coding tools perform similarly or slightly better to API-only models, hinting that code-specific models perform better when tasked to output data as code.

Since the Measurement F1 comprises 50\% of the overall F1, these low performances are highly detrimental. However, as demonstrated in  \autoref{extract_individual_properties}, we show that when models are tasked to solely extract individual measurements, they achieve substantially higher F1 scores, reaching up to 0.95. Furthermore, \autoref{appendix:assembing_graph_from_ground_truth} shows that when LLMs are tasked with organizing ground truth measurements into materials, the experiment extraction F1 reaches a high of 0.92.

By reviewing the model's outputs, we can pinpoint the deficiency in the Config F1 score to the omission of microstructure information. A core weakness is that models overlook information that is provided indirectly. For example, in \cite{shi2019enhanced}, the paper indirectly states that the DPHL740 and DPHL660 samples contain intragranular B2 grains with this statement: `[such] structural characteristics were also seen in the other two DPHL HEAs' \cite{shi2019enhanced}. This extraction is deceptively challenging because the paper made four DPHL materials instead of three, but the LLM needs to understand that the 660, 700, and 740 are the main samples, so that the ``other two" samples refer to DPHL660 and DPHL740. Gemini 3.1 Pro Preview failed to extract this indirect information.

Extracting process conditions is easier than extracting other pieces of information. A likely explanation is that most alloy papers include a `Methods' section that describes the process conditions in a single contiguous location, reducing the challenge of extracting those conditions. These findings suggest that future efforts should prioritize improving the extraction of configurations and measurement values.

KnowMat2 performs experiment extraction by repeatedly (three times) using multiple LLMs to first extract measurements and then validate for correctness. Despite its high cost, its low `Material' F1 score weakens its performance because it extracts data in a composition-centric manner, drastically reducing the number of extracted and matched measurements. Additionally, it has a low `Config' F1 score because it is designed to extract a flat list of measurements rather than microstructure. When evaluated with GPT-5 (high reasoning), a less descriptive LLM, KnowMat2 demonstrates that canonicalization is easier when performed during extraction, not after. Consider the extraction of \cite{jia2019novel}, which correctly extracts the compressive yield strength for $\mathrm{Al}_{19.9}\mathrm{Li}_{30}\mathrm{Mg}_{35}\mathrm{Si}_{10}\mathrm{Ca}_5\mathrm{Y}_{0.1}$ as `Yield Strength'. However, the post-processing LLM maps it to the canonical value of tensile yield strength because it lacks sufficient understanding to recognize that it was measured under compression. This example shows that canonicalization should occur concurrently with extraction, as additional context can help determine the correct canonical value.

To assess whether the number of attempts is correlated with performance, we computed the Pearson correlation coefficient between the overall F1 score and the number of attempts, yielding a value of $-0.2320$, indicating that fewer attempts lead to higher accuracy. However, the test is inconclusive because this correlation is weak, especially because the number of attempts is model dependent. For example, the Gemini models often take the most attempts but yield the highest overall scores, contradicting this coefficient.

\subsection*{Extracting Synthesis Steps Per-Material}

To illustrate why composition-centric extraction is insufficient, we compare LeMat-Synth \cite{lederbauer2025lemat} with frontier LLMs to extract synthesis conditions (benchmarking details are in \autoref{appendix:lematsynth}). The results in \autoref{tab:process_conditions_extraction} show that even when the task is restricted to synthesis processing steps, LeMat-Synth's composition-centric approach hinders the model's performance. One failure occurred while extracting \cite{haas2019microstructure}, where it assigned two annealing steps back-to-back for the same material: one at 1220 $^{\circ}$C and another at 950 $^{\circ}$C, thereby failing to recognize that the annealing steps correspond to different materials. Although LeMat-Synth’s General Synthesis Ontology is theoretically capable of extracting per-material synthesis conditions, LitXAlloy reveals that their prompts do not elicit this capability. This result reinforces the need for extraction at the material level.

\begin{table}[h]
\centering
\begin{tabular}{lc}
\toprule
\textbf{Model} & \textbf{F1} \\
\midrule
LeMat-Synth (Opus 4.6) & 0.58 $\pm$ 0.00 \\
KnowMat2 (GPT-5.2 High) & 0.67 $\pm$ 0.14 \\
\cmidrule(lr){1-2}
Claude Haiku 4.5 & 0.84 $\pm$ 0.06 \\
GPT-5 Mini Medium & 0.84 $\pm$ 0.07 \\
Gemini 3 Flash & 0.86 $\pm$ 0.02 \\
\cmidrule(lr){1-2}
Claude Opus 4.6 & 0.86 $\pm$ 0.02 \\
GPT-5.2 High & 0.85 $\pm$ 0.03 \\
Gemini 3.1 Pro & 0.83 $\pm$ 0.08 \\
\cmidrule(lr){1-2}
Claude Code (Opus 4.6) & \textbf{0.88 $\pm$ 0.01} \\
Codex (GPT-5.2 Codex High) & 0.82 $\pm$ 0.03 \\
Gemini CLI (3.1 Pro Preview) & 0.84 $\pm$ 0.04 \\
\bottomrule
\end{tabular}
\vspace{6pt}
\caption{Model performance on the process condition extraction task. LeMat-Synth was also evaluated with Gemini 3.1 Pro Preview (F1=0.50) and GPT-5.2 High (F1=0.49).}
\label{tab:process_conditions_extraction}
\end{table}

\subsection*{Extracting Compositions} \label{section:extracting_compositions}

Experimentally synthesized compositions are the most important measurements to index, as a material's composition strongly affects its properties. We benchmark frontier LLMs under two scenarios: one in which it outputs a string expressed in atomic percent, parsed by Pymatgen \cite{ong2013python}, and another in which it returns a Pymatgen Composition object. To assist LLMs in extracting compositions, the second scenario provides helper functions to normalize compositions, such as the one presented in \autoref{fig:comp_wt_additions_helper}, which simplifies complex logic into a single call.

\begin{table}[ht]
    \centering
    \begin{tabular}{lcc}
        \toprule
        \textbf{Model} & \textbf{F1} & \textbf{Total Cost} \\
        \midrule
        \multicolumn{3}{l}{\textit{Output string}} \\
        \cmidrule(lr){1-3}
        Claude Haiku 4.5        & $0.78 \pm 0.15$ & 0.28 \\
        GPT-5 Mini Medium       & $\mathbf{0.97 \pm 0.00}$ & 0.50 \\
        Gemini 3 Flash Preview         & $0.94 \pm 0.12$ & 0.54 \\
        \cmidrule(lr){1-3}
        Claude Opus 4.6         & $0.91 \pm 0.09$ & 1.38 \\
        GPT-5.2 High            & $\mathbf{0.97 \pm 0.00}$ & 0.96 \\
        Gemini 3.1 Pro Preview         & $\mathbf{0.97 \pm 0.00}$ & 0.97 \\
        \midrule
        \multicolumn{3}{l}{\textit{Output code (with helper functions)}} \\
        \cmidrule(lr){1-3}
        Claude Haiku 4.5        & $0.77 \pm 0.14$ & 0.28 \\
        GPT-5 Mini Medium       & $0.98 \pm 0.02$ & 0.48 \\
        Gemini 3 Flash Preview         & $\mathbf{0.99 \pm 0.01}$ & 0.50 \\
        \cmidrule(lr){1-3}
        Claude Opus 4.6         & $0.97 \pm 0.00$ & 1.49 \\
        GPT-5.2 High            & $0.97 \pm 0.00$ & 0.64 \\
        Gemini 3.1 Pro Preview         & $\mathbf{0.98 \pm 0.01}$ & 0.96 \\
        \bottomrule
    \end{tabular}
    \vspace{10pt}
    \caption{Model performance on the composition extraction task.}
    \label{tab:extract_composition}
\end{table}

As seen in \autoref{tab:extract_composition}, even without the assistance of code, models tasked with outputting composition strings are able to parse compositions quite well. Overall, outputting the composition as code (with helper function assistance) slightly improves the extraction, and models that reason longer, such as Claude Opus 4.6, experience significant gains. The prompt for this task is in \autoref{appendix:comp_extract_prompt}.

\subsection*{Extracting Individual Properties}\label{extract_individual_properties}

Given the cost of extracting all experimental measurements, a cheaper indexing strategy is to track the measurements for a single property. For example, it is useful to know which papers made a material with a measured Vickers hardness greater than 450 HV. Accordingly, we evaluate the LLMs to extract these measurements for five mechanical properties. The prompt for this task is found in \autoref{appendix:single_property_extraction_prompt}.

\begin{table}[h]
\centering
\begin{tabular}{lccc}
\toprule
\textbf{Property} & \textbf{Prec.} & \textbf{Rec.} & \textbf{F1} \\
\midrule
Ultimate Strength (T) & 0.83 & 1.00 & $0.91 \pm 0.00$ \\
Ultimate Strength (C) & 0.84 & 1.00 & $0.91 \pm 0.00$ \\
Fracture Strain (T)   & 0.85 & 0.96 & $0.90 \pm 0.08$ \\
Fracture Strain (C)   & 0.82 & 0.95 & $0.88 \pm 0.00$ \\
Vickers Hardness      & 0.96 & 0.93 & $0.95 \pm 0.04$ \\
\bottomrule
\end{tabular}
\vspace{8pt}
\caption{Performance of Gemini 3.1 Pro Preview on the single property extraction task. T = Tensile test. C = Compression test.}
\label{tab:single_prop}
\end{table}

\autoref{tab:single_prop} shows that LLMs can extract individual properties and basic mechanical properties well. The F1 for extracting individual properties is higher than the Measurement F1 for the experiment extraction task in \autoref{tab:zero_shot_code}, suggesting that without focusing on extracting other values, such as processing conditions and parent-child sample hierarchy, the model can devote its attention to extracting the measurements. These findings indicate that extracting per-property measurements is a viable indexing strategy.

\subsection*{Limitations}

While LitXBench captures many measurements per paper, this benchmark does not extract information from images or figures. We defer to other works such as PolyCompChartIE \cite{circiinformation} or LeMat-Synth \cite{lederbauer2025lemat}, which has a dedicated figure extraction benchmark using WebPlotDigitizer \cite{WebPlotDigitizer}. Moreover, the extracted values may contain non-obvious errors because the authors of the papers included in this benchmark were not consulted. This is acceptable as the open-source nature of this benchmark invites corrections. Furthermore, LitXBench currently does not track significant figures for numerical values because during code compilation, numbers expressed in Python lose their significant figures. Future work could preserve these values by enforcing numbers to be extracted as strings.

The datamodel in LitXBench is currently tailored to alloys. Therefore, extending the benchmark to additional material classes requires updates in three areas. First, the canonical values, which are currently specialized for alloys, must be adapted to the target material class. Second, new structured Measurement classes are required to capture complex, material-specific measurements, similar to the \texttt{Configuration} class discussed in \autoref{appendix:custom_measurement_classes}. Lastly, material-class-specific validation rules are needed to catch semantic extraction issues, similar to how casting steps must follow melting steps for alloys. Given the complexity of the adjustments needed to extend LitXBench to additional material classes, extraction is best performed on a per-material-class basis.

\section{Conclusion}

LitXBench is a benchmark for evaluating methods on the experiment extraction task. This work shows that measurements grouped by composition can obscure the material described. Instead, models should extract the process conditions and measured properties in a material-centric manner. In addition, we highlight the importance of mapping extracted categories to canonical values to avoid ambiguity. Furthermore, LitXBench argues that code-based extraction benchmarks provide validation benefits for annotators and are much more auditable. The benchmark shows that frontier LLMs are proficient at extracting synthesized materials, but stronger models or new methods are needed to accurately extract experiments in their entirety. We find that extracting compositions is quite easy, but extracting synthesis conditions and individual measurements is less reliable with current models. By formalizing the experiment extraction problem, LitXBench will motivate robust experiment extraction pipelines, enabling the creation of literature-derived datasets for materials science discovery.

\section*{Code Availability}

The code used for this work is available at \url{https://github.com/Radical-AI/litxbench}. The benchmark leaderboard and documentation are available at \url{https://radical-ai.github.io/litxbench}.

\section*{Acknowledgements}

This work was supported by Radical AI, Inc. The authors thank Yuval Krimer, Michael Pavel, Keithen Orson, and Rhys Goodall for their assistance with double-checking the extracted values in the benchmark. Conversations with Rhys Goodall on ontologies were particularly useful. We also appreciate Rhys Goodall, Stefano Falletta, Luke Pereira, Delia McGrath, and Yuri Sanspeur for their extremely constructive reviews of this manuscript.

\bibliography{litxbench}
\bibliographystyle{icml2026}

\begin{figure*}[h]
\centering
\vspace{2pt}
\includegraphics[width=0.8\textwidth]{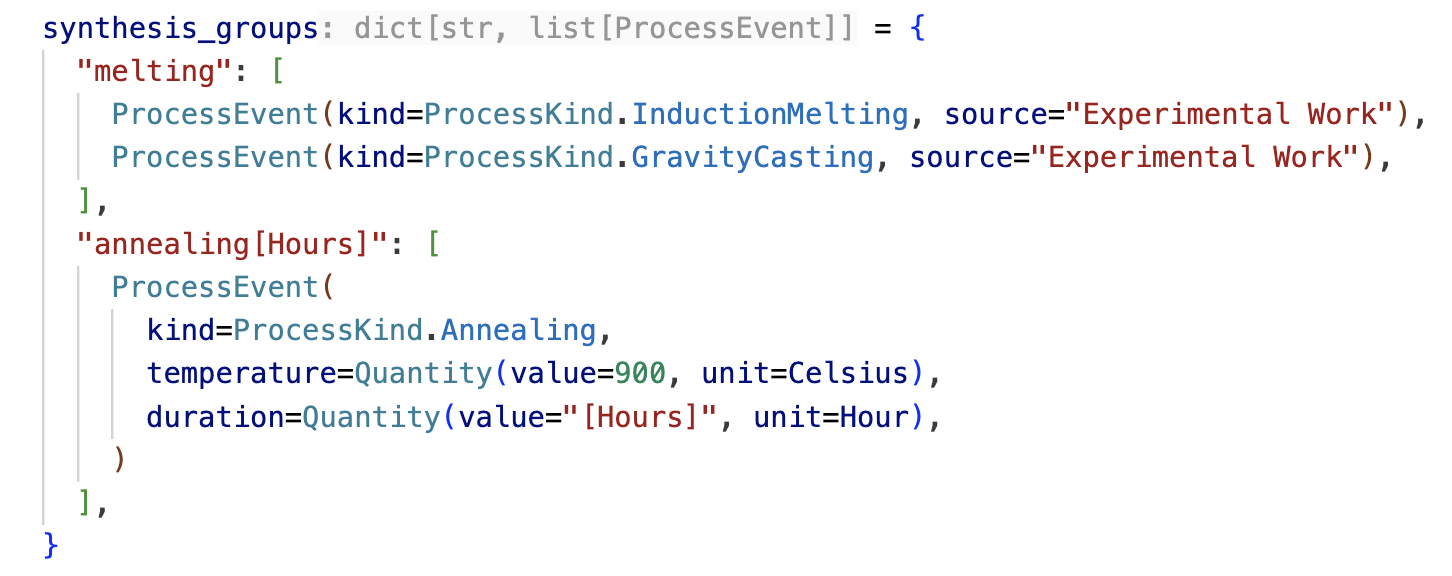}
\caption{Definition of each Synthesis Group. Each material defines which group of synthesis events it undergoes through the arrow notation \texttt{group1$\rightarrow$group2}. Groups that accept parameters (such as \texttt{Hours}) enable annotators to reuse synthesis groups across materials that differ by slight experimental parameters.}
\label{fig:synthesis_groups_example}
\end{figure*}

\newpage
\appendix

\section{Benchmark Details}\label{appendix:benchmark_details}

\subsection*{Benchmark Statistics}

There are 101 target materials in the dataset and 68 unique compositions. To highlight the importance of distinguishing materials with the same composition, of the 19 papers in this benchmark, 8 contain duplicate compositions, totaling 12 duplicate compositions overall. Furthermore, across 6 papers, 26 materials are derived from other materials in the dataset, highlighting the importance of preserving the graph-like dependence between materials when extracting them. LitXAlloy contains only experimental and experimentally-derived measurements. Computational measurements, such as those predicted by Thermo-Calc \cite{andersson2002thermo}, are not included.

\subsection*{Benchmark Quality}\label{appendix:benchmark_quality}

To ensure consistency, a single annotator performed the extraction for all LitXAlloy papers. The extracted values were also compared with those in MPEA to safeguard against missing values in LitXAlloy. Since humans are fallible, LLMs were employed to double-check and catch extraction mistakes. An estimated 1.1 billion tokens were spent using Opus 4.5 and 4.6 in Claude Code to catch errors. Additional scripts using Gemini 3.1 Pro Preview were used to identify errors and property-specific measurements. These machine-assisted techniques caught dozens of errors that the annotators missed, despite having thoroughly read each paper multiple times. All LLM-suggested corrections were heavily scrutinized by humans before LitXAlloy was updated. Despite best efforts, LitXAlloy likely contains errors, as a few were discovered during our experiments, prompting the re-evaluation of all models upon correction. Since errors were found in the benchmark after many rounds of corrections, human-annotated datasets without sufficient proof or annotations of correctness likely contain numerous errors.

\subsection*{Comparison with MPEA}

For the 18 papers that intersect with the MPEA dataset, our benchmark demonstrates significantly higher data density, with an average of 74.8 extracted measurements per paper, compared to 33.4 in MPEA. Within this overlapping subset, LitXAlloy contains an additional 745 values. Since LitXAlloy's annotation scope is much smaller than MPEA's, its data quality is likely higher. Notably, only $86.9\%$ of the phases and $87.40\%$ of the measured values match between LitXAlloy and MPEA, indicating that the MPEA dataset may have noticeable inaccuracies.

To ensure a fair comparison with MPEA, we define a value as a unique number/string. Hence, the temperature and pressure at which a measurement was taken are counted as separate values. Another concern is that MPEA has a lower ontological fidelity for measurement properties. For example, it maps compressive and tensile fracture strain to the same property. Thus, the comparison process only considers the numeric value of the measurement and ignores the measurement kind, thereby boosting the match rate between LitXAlloy and MPEA. To improve consistency in the comparison, purely computational properties, such as those predicted by Thermo-Calc, were excluded. For simplicity, we ignore the temperature when the measurement is made. Since the recorded microstructure in MPEA is not as detailed as that in LitXAlloy, phases in MPEA classified as `other' and not FCC/BCC are considered to match the target configuration.

\section{LitXAlloy Data Models} \label{appendix:data_models}
This section outlines the data models that describe the extracted materials. Some measurement methods, such as tensile tests, are destructive and therefore require multiple samples to be made. For readability, the extraction target is a list of materials, rather than all individually synthesized samples. Each material is defined as the consolidation of samples with identical processing events.

\begin{figure}[t]
\centering
\vspace{2pt}
\includegraphics[width=0.4\textwidth]{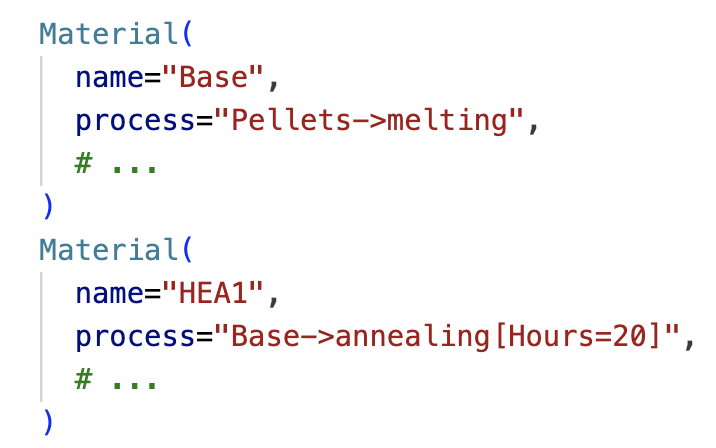}
\caption{Examples of process definitions for two materials. Upper: A material that is created from the \texttt{Pellets} raw\_material. Lower: A material that is derived from the \texttt{Base} material rather than from raw input materials. This material also specifies its annealing duration to the synthesis group.}
\label{fig:synthesis_process_example}
\end{figure}

The primary goal of defining process conditions is to express them in a straightforward, strict temporal order to reduce ambiguity, in accordance with the recommendation in \cite{kim2019distilling}. Process conditions are first defined for each experiment, as seen in \autoref{fig:synthesis_groups_example}. Next, each material specifies its synthesis process, as illustrated in \autoref{fig:synthesis_process_example}. Raw input materials are followed by the arrow notation \texttt{->}, which denotes subsequent processing events. Process events are denoted using this custom syntax because it is much more compact and readable than using a list of Python objects. To reduce annotator error, processing events may optionally accept template parameters, allowing different materials to reference the same synthesis group with different parameters and thereby maximizing code reuse.

Since multiple base materials can be combined to form downstream materials \cite{xu2018microstructure}, experiments can therefore be conceptualized as a directed acyclic graph (DAG). This data model empowers child materials to specify their parents. This is more flexible than having parent materials specify child materials, since child materials can have multiple parents. In addition, template variables have the flexibility to accept other materials as input. Although materials are defined in a graph-like manner, processing events extracted linearly rather than in a DAG-like relationship may affect performance. Various extraction formats are explored in \autoref{appendix:graph_repr}.

\subsection*{Storing Measurements}\label{appendix:custom_measurement_classes}

Most experimental measurements are simple and can be recorded with an instance of the \texttt{Measurement} class. As seen in \autoref{fig:sample_code_example}, Measurements contain a \texttt{MeasurementKind} with a \texttt{value} and an optional Pint \cite{grecco_pint} \texttt{unit}. Measurements also contain an optional uncertainty value. Uncertainty estimates are inconsistent between papers, as authors often use different methods to quantify uncertainty. All measurements contain a \texttt{MeasurementMethod} enum because the method used to take a measurement is a strong indicator of data quality.

The measurement class is sufficient to index simple datapoints, but if required, additional information may be stored in its \texttt{description} field. However, some measurements, such as lattice parameters, are complex and are described by many more attributes. In these cases, custom dataclasses are needed to store this information. \texttt{LatticeMeasurement} and \texttt{GlobalLatticeParam} store lattice information using the \texttt{Lattice} class in \texttt{Pymatgen}. Since the extraction format is code, the constructors and validation logic of the \texttt{Lattice} class are automatically inherited. The \texttt{CompMeasurement} class represents composition using Pymatgen's \texttt{Composition} class. This is advantageous as \texttt{Composition} contains functions to normalize compositions expressed in atomic percent or weight percent.

To map categories of extracted values to canonical values, LitXAlloy provides a manually curated set of enums such as \texttt{MeasurementKind} and \texttt{RawMaterialKind}. These enums are not organized hierarchically as in most ontologies \cite{del2024elementary}, since declaring them as a set of flattened enums is sufficient for indexing purposes. Discretion is necessary to identify the enums best suited for indexing applications before applying the LitXAlloy enums to other material classes. Canonical values are best represented as enums, as they are disjoint. In addition, extracting fields as enums (rather than strings) offers a compile-time guarantee that the field is a valid canonical category.

The \texttt{normalize} function is used to map a string that appears in the text to the canonical value for that category. This serves as documentation for the mapping between the string in the paper and the correct canonical value. This function also teaches mislabeled properties in the original paper. For example, the normalize function is used in the extracted measurements of \cite{sanchez2019design} to note that the `plastic strain' measurements are in reality `compressive fracture strain' measurements.

\section{Cost Functions for Hungarian Scoring}
\label{appendix:scoring_cost}

\subsection*{Overall F1 Score Calculation} \label{appendix:overall_f1}

The overall F1 score for LitXBench is composed of individual F1 scores for the extraction of measurements, process conditions, materials, and microstructure configurations. The overall F1 score is a weighted sum of these individual extractions, where the weights for each contribution are in \autoref{tab:overall_score_weights}. Extracting the measurements is the most important, so it makes up half of the overall F1 score. The configuration and material scores are weighted the least because they primarily ensure that the number of materials and microstructure configurations is correctly identified.

\begin{table}[t]
\centering
\renewcommand{\arraystretch}{1.2}
\resizebox{\columnwidth}{!}{%
\begin{tabular}{llc}
\toprule
\textbf{Category} & \textbf{Description} & \textbf{Weight} ($W$) \\ 
\midrule
Measurements  & Measurement values     & 0.50 \\
Process       & Process condition      & 0.20 \\
Material      & Set of materials       & 0.15 \\
Configuration & Set of microstructure  & 0.15 \\
\bottomrule
\end{tabular}%
}
\vspace{2pt}
\caption{Weighted scoring scheme for the overall F1 score.}
\label{tab:overall_score_weights}
\end{table}

The Measurements score determines the similarity between the extracted and target measurements, regardless of which material they are for. To determine matches, we first ensure that the measurements are of the same kind. Next, the value and unit of the measurement is compared. Lastly, the uncertainty quantifiers (e.g. \textbf{$<=$}; weight=1), the temperature (weight=2), and the pressure (weight=2) of the measurements are taken into account.

Scoring process conditions are unique because they must be extracted in the correct order. Therefore, we use the Levenshtein distance algorithm \cite{levenshtein1966binary} to determine the Hungarian cost.

The Material score is used to evaluate whether the correct amount of materials is extracted. This is calculated by first applying the Hungarian algorithm on both sets of extracted and target materials. The resulting pairwise matchings affect the precision and recall for the Material score.

Similar to materials, the Configurations score is used to evaluate whether the correct amount of configurations are extracted. This is the most involved score because configurations of a material can form a tree. For example, the paper may reference that ``Spot A" and ``Spot B" are different sub-configurations of the BCC phase. As a result, configurations can form a dependency hierarchy referencing other configurations they are within. We solve this problem similarly to how we score materials: we apply the Hungarian Algorithm to best match the measurements within configurations (ignoring hierarchical dependencies between configurations). To account for the hierarchical dependencies, we then identify graph Markov equivalence. An equivalence score is computed to determine whether the parent node in each extracted configuration matches the parent node of the target configuration in its respective graph. By applying the Hungarian algorithm first, this scoring scheme prioritizes matching measurements over graph isomorphism, as extracting the correct measurements is more important than the relationships between configurations.

\section{JSON Output Extraction Format} \label{appendix:json_extraction}

We evaluated if the LLMs would perform better if they outputted experiments as JSON rather than in code. Once extracted, JSON outputs were converted to the code schema outlined in \autoref{appendix:data_models} before evaluation, as usual. Validation errors were properly converted to omit references to code and instead flag the error in the corresponding JSON output. The prompt for this experiment is found in \autoref{appendix:json_extraction_format_prompt}.

\begin{table}[h]
\centering
\begin{tabular}{lcccc}
\toprule
& \multicolumn{2}{c}{\textbf{Output JSON}} & \multicolumn{2}{c}{\textbf{Output Code}} \\
\cmidrule(lr){2-3} \cmidrule(lr){4-5}
\textbf{Model} & F1 & Attempts & F1 & Attempts \\
\midrule
Claude Haiku 4.5  & 0.63 & 3.6 & \textbf{0.65} & 2.2 \\
GPT-5 Mini Med.   & 0.65 & 3.5 & \textbf{0.68} & 2.5 \\
Gemini 3 Flash    & \textbf{0.76} & 3.1 & 0.74 & 2.6 \\
\cmidrule(lr){1-5}
Claude Opus 4.6   & 0.72 & 2.2 & \textbf{0.72} & 1.5 \\
GPT-5.2 High      & 0.69 & 2.0 & \textbf{0.73} & 1.5 \\
Gemini 3.1 Pro    & 0.76 & 2.2 & \textbf{0.77} & 1.5 \\
\bottomrule
\end{tabular}
\vspace{2pt}
\caption{Extraction performance of code vs JSON output.}
\label{tab:json_performance}
\end{table}

As seen in \autoref{tab:json_performance}, the JSON output format performed slightly worse (a lower F1 score of up to 0.03), likely because frontier LLMs spend more time being post-trained for writing code rather than outputting JSON. However, this difference is considered insignificant.

\section{LeMat-Synth Experimental Setup} \label{appendix:lematsynth}

Since LeMatSynth only extracts synthesis conditions, we only benchmarked this capability on LitXBench. During extraction, LeMat-Synth used the transcribed text in LitXBench only to test the extraction pipeline and eliminate transcription inconsistencies. The \texttt{synthesis\_extraction} and \texttt{material\_extraction} LLMs both used Opus 4.6. Because the pipeline lacks access to our \texttt{ProcessKind} enum during extraction, we map each of LeMatSynth's \texttt{action}s to the corresponding \texttt{ProcessKind}. We were generous in our conversion as generic actions such as `heat' were matched to more specific \texttt{ProcessKind} types. Extraneous actions such as `add', `flip', `remelt', and `stack' were removed before evaluation because these steps are a stylistic difference between LitXBench and LeMat-Synth and would unjustly hurt LeMat-Synth's score. Once LeMat-Synth's set of \texttt{ProcessKind} events has been created, they are matched to the ground truth process conditions using dynamic programming before the F1 score is calculated.

\section{KnowMat2 Experimental Setup} \label{appendix:knowmat2}

Similar to LeMat-Synth, during extraction, KnowMat2 used the transcribed text in LitXBench only to test the extraction pipeline and eliminate transcription inconsistencies. The extraction, manager, and evaluation models all used GPT-5.2 high, as these models performed the hardest extraction tasks. The subfield and flagging models used the default GPT-5 Mini model. The properties.json file was modified to specify the canonical names used by LitXBench. During evaluation, the \texttt{standard\_property\_name} was first checked to map the property to an \texttt{AlloyMeasurementKind} or \texttt{PhaseMeasurementKind}. If that failed, fuzzy matching logic was performed on \texttt{property\_name} instead. \texttt{ProcessKind} were identified using fuzzy string matching, as KnowMat2 specifies all process conditions in a single extracted string.

\section{Graph vs List Extraction Format}\label{appendix:graph_repr}

To isolate the effect of the graph schema outlined by LitXBench, we perform two experiments. The first experiment provides a single paper's text to an LLM along with all extracted target materials. The twist is that the extracted target materials are anonymized as all measurement values are redacted. The LLM is tasked to identify the set of anonymized target materials that best fit the paper. We ask the LLMs to do this twice: once with the anonymized materials laid out flat, and once with sample relationships included in the redacted extraction targets. The hypothesis was that the structural graph information would help the models better identify the experiments. However, from \autoref{table:match_graph_vs_flat}, it is clear that the graph structure does not help the models discern the proper experimental values for a paper. The prompts for this experiment are found in \autoref{appendix:match_paper_to_correct_experiment}.

\begin{table}[H]
\centering
\begin{tabular}{l c c}
\toprule
\textbf{Model} & \textbf{Match Graph} & \textbf{Match Flat} \\
\midrule
Claude Haiku 4.5         & \textbf{0.32} & \textbf{0.32} \\
GPT-5 Mini Medium        & 0.42          & \textbf{0.47} \\
Gemini 3 Flash Preview          & \textbf{0.47} & 0.42          \\
\cmidrule(lr){1-3}
Claude Opus 4.6          & \textbf{0.58} & \textbf{0.58} \\
GPT-5.2 High             & \textbf{0.53} & 0.47          \\
Gemini 3.1 Pro Preview          & \textbf{0.79} & \textbf{0.79} \\
\bottomrule
\end{tabular}
\vspace{4pt}
\caption{Matching accuracy of anonymized material extractions.}
\label{table:match_graph_vs_flat}
\end{table}

The second experiment determines whether extracting the materials in a graph-like or flat manner is best. A flat extraction is when a child material's dependency on its base materials is omitted. Thus, all process chains must start from raw materials. The results are outlined in \autoref{table:extraction_graph_vs_flat}. They show that graph extractions yield results similar to flat extractions. The differences between the normal extraction task and the flat extraction task are that all process conditions must start from a \texttt{raw\_material} rather than possibly starting from another base material. The instruction prompts in \autoref{appendix:example_extraction_prompt} were modified for this experiment.

\begin{table}[h]
\centering
\begin{tabular}{l c c}
\toprule
\textbf{Model} & \textbf{Extract Graph} & \textbf{Extract Flat} \\
\midrule
Claude Haiku 4.5         & 0.68 & \textbf{0.69} \\
GPT-5 Mini Medium        & 0.68 & \textbf{0.70} \\
Gemini 3 Flash Preview          & \textbf{0.77} & 0.73 \\
\cmidrule(lr){1-3}
Claude Opus 4.6          & \textbf{0.74} & 0.74 \\
GPT-5.2 High             & 0.69          & \textbf{0.71} \\
Gemini 3.1 Pro Preview          & 0.76          & \textbf{0.78} \\
\bottomrule
\end{tabular}
\vspace{4pt}
\caption{Updated Extraction format F1 scores.}
\label{table:extraction_graph_vs_flat}
\end{table}

The results in \autoref{table:match_graph_vs_flat} and \autoref{table:extraction_graph_vs_flat} indicate that no clear extraction format is advantageous. However, benchmarks such as LitXBench should still be written in a graph-like manner because it keeps experimental information in a compact form, reducing the chance of errors during edits. In addition, graphs can be converted to a list of items, but not the other way around (since graphs contain dependency information). These reasons advocate for future benchmarks to be written in a graph-like fashion when the data has graph-like dependencies.

\section{Assembling Material Graph from Ground Truth Values}\label{appendix:assembing_graph_from_ground_truth}

To isolate the model's ability to organize and cluster data into the correct materials, an experiment was conducted in which the ground-truth processing conditions and measurements were presented to the LLM in a disordered manner. The LLM's task is to convert this set of numbers into a set of \texttt{Materials} when given the paper. From the results in \autoref{tab:assemble_graph_from_ground_truth}, we clearly see that the results are better than those of the pure-extraction task in \autoref{tab:zero_shot_code}. This is expected, as the model does not need to identify the important values from the paper; it only needs to reason about the material each value belongs to.

\begin{table}[h]
\centering
\begin{tabular}{lcccc}
\toprule
\textbf{Model} & \textbf{Prec.} & \textbf{Rec.} & \textbf{F1} & \textbf{Attempts} \\
\midrule
Claude Haiku 4.5 & 0.75 & 0.78 & 0.75 & 1.95 \\
GPT-5 Mini Med.  & 0.73 & 0.78 & 0.75 & 2.21 \\
Gemini 3 Flash   & 0.89 & 0.88 & 0.88 & 1.58 \\
\cmidrule(lr){1-5}
GPT-5.2 High     & 0.83 & 0.89 & 0.85 & 1.53 \\
Claude Opus 4.6  & 0.86 & 0.89 & 0.87 & \textbf{1.42} \\
Gemini 3.1 Pro   & \textbf{0.92} & \textbf{0.94} & \textbf{0.92} & 1.79 \\
\bottomrule
\end{tabular}
\vspace{6pt}
\caption{Assembling the material graph from ground truth values.}
\label{tab:assemble_graph_from_ground_truth}
\end{table}
\section{Figures Highlight Textual Errors}\label{appendix:figure_error}

Including figures may present more information, which can reveal mistakes in the paper text. For example, in Table 3 of \cite{chen2014microstructures}, measurements are labeled as `fracture strain'. However, when looking for those points on the stress-strain curve, these measurements actually refer to the `strain at the ultimate point'. Since LitXBench only considers textual information, the current ground truth for those measurements is `fracture strain' as the body text cannot reveal this error. However, when this benchmark includes figures, the target measurement will be updated to `strain at the ultimate point'.

\onecolumn
\section{Evaluation Prompts}

\subsection*{Experiment Extraction Prompts}\label{appendix:experiment_extraction_prompts}\label{appendix:example_extraction_prompt}

Here are the prompts for the experiment extraction task described in \autoref{section:experiment_setup}. We provide all models with an example of the experiments they should extract. To prevent data leakage, we ensured that the materials shown in the prompt did not exactly match the ground-truth materials.

\begin{prompt}[]
[
  Experiment(
    raw_materials={
      "elements": RawMaterial(
        kind=RawMaterialKind.Ingot,
        description="...",
        source="...",
      )
    },
    synthesis_groups={
      "creation": [
        ProcessEvent(kind=ProcessKind.ArcMelting, description="Remelted 5 times to promote homogeneity, flipped pellet between each remelt", source="..."),
        ProcessEvent(kind=ProcessKind.AsCast, description="...", source="..."),
        ProcessEvent(kind=ProcessKind.Homogenization, temperature=Quantity(value=1200, unit=Celsius), duration=Quantity(value=24, unit=ureg.hour), source="..."),
        ProcessEvent(kind=ProcessKind.WaterQuenching, source="..."),
      ],
      "annealing[Temp]": [
        ProcessEvent(
          kind=ProcessKind.Annealing,
          temperature=Quantity(value="[Temp]", unit=Celsius),
          source="...",
        ),
      ],
    },
    descriptions=[
      AlloyDescriptionGroup(
        kinds=[AlloyMeasurementKind.vickers_hardness],
        method=MeasurementMethod.VickersHardnessTest,
        desc="The microhardness (HV) of the alloys was measured with a Vickers hardness tester under a load of 500 gf",
      ),
      AlloyDescriptionGroup(
        kinds=[ProcessKind.Grinding],
        desc="performed on a grinding wheel",
      ),
    ],
    output_materials=[
      Material(
        process="elements->creation",
        name="materialA",
        measurements=[
          CompositionMeasurement("MgFeNi"),
          Measurement(
            kind=AlloyMeasurementKind.vickers_hardness,
            value=321.0,
            unit=HV,
            uncertainty=7.0,
            source="...",
          ),
          # When a value is explicitly a mean/average:
          Measurement(
            kind=AlloyMeasurementKind.yield_strength_tension,
            value=450.0,
            unit=MegaPascal,
            measurement_statistic=MeasurementStatistic.mean,
            source="...",
          ),
          # When a value is reported as a range (e.g. "5-50 um"):
          *Measurement.group_measurements(
            kind=PhaseMeasurementKind.grain_size,
            unit=Micrometer,
            values=[
              CoreMeasurementValue(statistic=MeasurementStatistic.lower, value=5),
              CoreMeasurementValue(statistic=MeasurementStatistic.upper, value=50),
            ],
          ),
          # XRD-determined lattice parameter and crystal structure:
          GlobalLatticeParam(
            lattice=LatticeMeasurement(Lattice.cubic(3.208)),
            struct=CrysStruct.BCC,
            source="...",
          ),
          # Microstructural feature (e.g. dendrite with EDS composition):
          Configuration(
            name="dendrite",
            tags={ConfigTag.Dendrite},
            measurements=[
              CompMeasurement(
                {"Mo": 27.6, "Nb": 25.0, "Ta": 31.5, "V": 15.9},
                method=MeasurementMethod.EDS,
                source="...",
              ),
            ],
          ),
          # Nested configuration (precipitate within a phase):
          Configuration(
            name="FCC matrix",
            struct=CrysStruct.FCC,
            tags={ConfigTag.Matrix},
            measurements=[
              Measurement(
                kind=PhaseMeasurementKind.grain_size,
                value="~0.71",
                unit=Micrometer,
              ),
            ],
          ),
          Configuration(
            name="B2 precipitates",
            struct=CrysStruct.B2,
            tags={ConfigTag.Precipitate, ConfigTag.Intragranular},
            within="FCC matrix",
            measurements=[
              *Measurement.group_measurements(
                kind=PhaseMeasurementKind.grain_size,
                unit=Nanometer,
                values=[
                  CoreMeasurementValue(statistic=MeasurementStatistic.lower, value=50),
                  CoreMeasurementValue(statistic=MeasurementStatistic.upper, value=180),
                ],
              ),
            ],
          ),
        ],
      ),
      Material(
        process="materialA->annealing[Temp=10]",
        name="materialB",
        measurements=[
          CompositionMeasurement("MgFeNi"),
          Measurement(
            kind=AlloyMeasurementKind.vickers_hardness,
            value=350.0,
            unit=HV,
            uncertainty=7.0,
            source="...",
          ),
        ],
      ),
    ],
  ),
]
\end{prompt}

In addition to the example, this prompt provides context for performing the extraction.

\begin{prompt}[]
How to write each Experiment and fields to populate:
1) `raw_materials` (required): map each initial input name (for example `"elements"` or `"powders"`) to `RawMaterial`.
 - Populate `kind` with `RawMaterialKind` (usually `Ingot`, `Powder`, or `Unspecified`).
 - Populate `description` and `source` whenever the paper states purity, supplier, or precursor details.
2) `synthesis_groups` (required): a dict of named synthesis stages to lists of `ProcessEvent`.
 - Use reusable stages and process variables when appropriate (for example `"annealing[Temp]"`).
 - Each `ProcessEvent` should include `kind` (a `ProcessKind` enum member), and include `temperature` (as `Quantity`, e.g. `Quantity(value=1200, unit=Celsius)`), `duration` (as `Quantity`, e.g. `Quantity(value=24, unit=ureg.hour)`), `description`, `source` when available.
 - Use `ProcessEvent.inputs` to declare which raw materials or named materials feed into a specific
  process event. This is useful when a raw material is introduced at a particular synthesis step
  rather than at the very start of the process. `inputs` can contain literal names or template
  variables. Examples:
  ```python
  # Literal: two raw materials are mixed together
  ProcessEvent(kind=ProcessKind.Mixing, inputs=["elements", "secondary_input_metal"], description="...")
  # Variable: the feedstock is parameterized so the group can be reused
  # synthesis group key: "mixing[Feedstock]"
  ProcessEvent(kind=ProcessKind.Mixing, inputs=["[Feedstock]"], description="...")
  # then in material process: "powder->milling->mixing[Feedstock=secondary_input_metal]"
  ```
 - Every raw material declared in `raw_materials` must be referenced somewhere - either as a
  comma-separated input in a `Material.process` string or via `ProcessEvent.inputs`.
 - `ProcessKind.ArcMelting` and `ProcessKind.InductionMelting` must always be immediately followed by a casting step (e.g. `ProcessKind.AsCast`, `ProcessKind.SuctionCasting`, `ProcessKind.DropCasting`, `ProcessKind.GravityCasting`, `ProcessKind.DirectionalSolidification`, or `ProcessKind.CastingUnspecified`).
 - IMPORTANT: Every synthesis group you define MUST be referenced by at least one material's `process` string.
  Do not define groups that are not used.
 - IMPORTANT: If a group name contains a template variable like `[Temp]` (e.g. `"annealing[Temp]"`),
  that exact placeholder string `"[Temp]"` MUST appear as a value in at least one `ProcessEvent` field
  within that group. For example: `temperature=Quantity(value="[Temp]", unit=Celsius)`.
  Do NOT put the actual number in the group definition - the actual value goes in the material's
  `process` string (e.g. `process="elements->annealing[Temp=700]"`).
 - Only include actual fabrication/processing steps in `synthesis_groups`.
  Do NOT put measurement methods, characterization techniques, or testing procedures here.
3) `output_materials` (required): list of `Material`.
 - Populate `Material.process` using dataset process notation such as
  `"elements->creation"` or `"base->annealing[Temp=700]->quenching"`.
 - The first segment (before the first `->`) is a comma-separated list of input raw materials
  or named materials. Use commas to combine multiple inputs: `"elements,reinforcement->mixing->sintering"`.
 - Include `Material.name` only if the paper names that material.
 - `Material.measurements` must include at least one `CompositionMeasurement`.
 - Add ALL reported property measurements with `Measurement(...)` using `AlloyMeasurementKind` members.
4) Measurements:
 - Use `Measurement(kind=AlloyMeasurementKind.<kind>, value=<number>, unit=<unit>)`.
 - If uncertainty is reported (e.g. "450 +- 20"), set `value=450.0` and `uncertainty=20.0`.
 - If temperature or pressure is tied to a measurement, set `temperature=Quantity(...)` or `pressure=Measurement(...)`.
 - Assume room temperature is ~23 C when the paper says "room temperature" without a number.
 - Pay attention to how the paper qualifies numeric values. Use the value field as follows:
  - Exact value: `value=50.0` (paper states "50 HV")
  - Approximate value: `value="~50"` (paper says "around 50 HV", "approximately 50 HV", "about 50 HV")
  - Greater than: `value=">50"` (paper says "above 50 HV", "greater than 50 HV", "exceeding 50 HV")
  - Less than: `value="<50"` (paper says "below 50 HV", "less than 50 HV", "under 50 HV")
  - Greater than or equal to: `value=">=50"` (paper says "at least 50 HV", "50 HV or more", "no less than 50 HV")
  - Less than or equal to: `value="<=50"` (paper says "at most 50 HV", "up to 50 HV", "no more than 50 HV")
  - Much greater than: `value=">>50"` (paper says "much greater than 50 HV", "far above 50 HV", "well above 50 HV")
  - Much less than: `value="<<50"` (paper says "much less than 50 HV", "far below 50 HV", "well below 50 HV")
 - When the paper explicitly states a value is a mean or average, set `measurement_statistic=MeasurementStatistic.mean`.
  Leave `measurement_statistic` as `None` (default) when the paper does not specify.
 - When a measurement is reported as a range (e.g. "5-50 um", "between 50 and 180 nm"), use `Measurement.group_measurements(...)`:
  ```python
  *Measurement.group_measurements(
    kind=PhaseMeasurementKind.grain_size,
    unit=Micrometer,
    values=[
      CoreMeasurementValue(statistic=MeasurementStatistic.lower, value=5),
      CoreMeasurementValue(statistic=MeasurementStatistic.upper, value=50),
    ],
  )
  ```
  This creates two linked Measurement objects (one lower, one upper) that share the same `group_id`.
  Note the `*` spread operator - use it inside a list to unpack the group into individual measurements.
5) GlobalLatticeParam (for XRD lattice parameters and crystal structure):
 - Use `GlobalLatticeParam` when the paper reports lattice parameters from XRD for the overall material.
 - `lattice`: wrap a pymatgen `Lattice` in `LatticeMeasurement(...)`. Required parameters depend on type:
  - `Lattice.cubic(a)` - requires `a`
  - `Lattice.hexagonal(a, c)` - requires `a` and `c`
  - `Lattice.tetragonal(a, c)` - requires `a` and `c`
  - `Lattice.orthorhombic(a, b, c)` - requires `a`, `b`, and `c`
 - `struct`: the crystal structure enum, e.g. `CrysStruct.BCC`, `CrysStruct.FCC`.
 - `phase_fraction`: optionally attach a `Quantity` for the phase fraction if reported.
 - `name`: optional name for the phase (e.g. `"BCC phase"`, `"sigma phase"`).
 - Include `source` noting where in the paper the data appears.
6) Configuration (for microstructural features):
 - Use `Configuration` to describe microstructural features like dendrites, precipitates, phases, lamellae,
  or regions of interest with distinct microstructure (e.g. a Cr-rich region, an interdendritic zone).
 - Do NOT use Configuration merely to record where on the bulk material a measurement was taken.
  If the paper says "hardness at the center region was 210 HV", that is a material-level measurement -
  put it in `Material.measurements`, not inside a Configuration named "center region".
 - `name`: identifies the feature (e.g. `"dendrite"`, `"FCC matrix"`, `"B2 precipitates"`).
 - `struct`: crystal structure if known (e.g. `CrysStruct.BCC`, `CrysStruct.FCC`, `CrysStruct.B2`).
 - `tags`: categorize the feature using `ConfigTag` members (e.g. `ConfigTag.Dendrite`, `(<ConfigTag.Precipitate: 'precipitate'>, <ConfigTag.Intragranular: 'intragranular'>)`).
 - `within`: reference the `name` of another Configuration in the same material to indicate nesting
  (e.g. precipitates within a matrix phase). The referenced Configuration must exist in the same material.
  All precipitates MUST have a `within` field referencing the configuration they are contained in.
 - `measurements`: list of measurements specific to this feature - typically `CompMeasurement` (EDS composition),
  `Measurement` (grain size, phase fraction, etc. using `PhaseMeasurementKind`), or `LatticeMeasurement`.
7) `descriptions` (optional): list of `AlloyDescriptionGroup` for recording contextual information about
 measurement methods and equipment, or process-related descriptions that apply to all materials.
 - Use this field for information about HOW measurements were performed (instruments, testing conditions,
  specimen dimensions, strain rates) and general descriptions of process steps (equipment details).
 - Do NOT put this information in `synthesis_groups` - synthesis groups are for the actual fabrication
  process chain, not for characterization or testing procedures.
 - `kinds`: list of `AlloyMeasurementKind`, `PhaseMeasurementKind`, `ProcessKind`, or `MeasurementMethod`
  members that this description applies to.
 - `method`: optional `MeasurementMethod` enum member (e.g. `MeasurementMethod.VickersHardnessTest`,
  `MeasurementMethod.CompressionTest`).
 - `desc`: free-text description of the method, equipment, or conditions.
 - Examples:
  ```python
  descriptions=[
    # Describes how hardness was measured
    AlloyDescriptionGroup(
      kinds=[AlloyMeasurementKind.vickers_hardness],
      method=MeasurementMethod.VickersHardnessTest,
      desc="Microhardness measured with Vickers hardness tester at 500 gf load for 15 s",
    ),
    # Describes compression testing conditions
    AlloyDescriptionGroup(
      kinds=[AlloyMeasurementKind.yield_strength_compression,
          AlloyMeasurementKind.fracture_strength_compression],
      method=MeasurementMethod.CompressionTest,
      desc="Compressive testing on MTS 809 machine at room temperature. Specimens: 3 mm x 6 mm.",
    ),
    # Describes a process step's equipment
    AlloyDescriptionGroup(
      kinds=[ProcessKind.Grinding],
      desc="performed on a grinding wheel",
    ),
  ]
  ```
\end{prompt}

This prompt defines the scope of the extraction. It instructs the LLM to limit extraction to experimental measurements for the samples the authors synthesized.

\begin{prompt}[]
We are only interested in materials the authors physically made in their lab.
Extract ALL properties reported for those materials - mechanical, thermal, physical, microstructural, etc.
Include properties derived from measurements (e.g. strain hardening exponent from a stress-strain curve, fracture toughness from a fracture test, Pugh ductility ratio from measured elastic constants).
Exclude properties that are purely computed from composition or thermodynamic databases (e.g. CALPHAD/ThermoCalc predictions, DFT-computed values, rule-of-mixtures estimates with no experimental validation).
\end{prompt}

Since canonicalization is performed at runtime (the benefits of which is described in more detail when discussing KnowMat2 in \autoref{section:results}), we provide the list of canonical values in a prompt. We also briefly mention the objects within the code execution environment at runtime.
\begin{prompt}[]
Available names in runtime
- Core classes: `Experiment`, `Material`, `RawMaterial`, `RawMaterialKind`, `ProcessEvent`, `ProcessKind`, `Measurement`, `CompositionMeasurement`
- Microstructure classes: `GlobalLatticeParam`, `LatticeMeasurement`, `Configuration`
- Lattice constructor: `Lattice` (from `pymatgen.core.lattice`) - e.g. `Lattice.cubic(a)`, `Lattice.hexagonal(a, c)`
- Measurement statistics: `MeasurementStatistic`, `CoreMeasurementValue`
- Enums: `AlloyMeasurementKind`, `PhaseMeasurementKind`, `ProcessKind`, `CrysStruct`, `ConfigTag`, `MeasurementMethod`, `ValueQualifier`
- `AlloyMeasurementKind` members: `vickers_hardness`, `berkovich_hardness`, `pugh_ductility_ratio`, `density`, `yield_strength_tension`, `ultimate_strain_tension`, `ultimate_tensile_strength`, `fracture_strain_tension`, `fracture_strength_tension`, `strain_hardening_exponent_tension`, `poissons_ratio_tension`, `fracture_energy_tension`, `true_stress_tension`, `yield_strength_compression`, `ultimate_strain_compression`, `ultimate_compressive_strength`, `fracture_strain_compression`, `fracture_strength_compression`, `strain_hardening_exponent_compression`, `poissons_ratio_compression`, `fracture_energy_compression`, `true_stress_compression`, `elastic_limit_compression`, `elastic_limit_tension`, `youngs_modulus`, `fracture_toughness`, `work_of_fracture`, `crystallite_size`, `lattice_strain`, `melting_point`, `solidus`, `liquidus`
- `PhaseMeasurementKind` members (for Configuration/GlobalLatticeParam measurements): `volume_fraction`, `length`, `grain_size`, `phase_size`
- `ProcessKind` members: `Mixing`, `MechanicalAlloying`, `PlanetaryMilling`, `GasAtomization`, `ArcMelting`, `InductionMelting`, `CastingUnspecified`, `AsCast`, `GravityCasting`, `DropCasting`, `SuctionCasting`, `DirectionalSolidification`, `SparkPlasmaSintering`, `HotPressingSintering`, `VacuumFurnace`, `Homogenization`, `Annealing`, `NonIsothermalAnnealing`, `IsothermalHolding`, `WaterQuenching`, `SolutionHeatTreatment`, `HotExtrusion`, `HotRolling`, `ColdRolling`, `CrossRolling`, `ColdForging`, `Press`, `FrictionStirProcessing`, `ElectricalDischargeMachining`, `Cut`, `Grinding`, `Polishing`, `Etching`, `AquaRegia`, `SandBlasting`, `Degreased`, `UltrasonicBath`, `AirDrying`
- `CrysStruct` members: `FCC`, `BCC`, `HCP`, `DHCP`, `Diamond`, `L12`, `L10`, `B2`, `D019`, `D03`, `Heusler`, `Rocksalt`, `Zincblende`, `C14`, `C15`, `Perovskite`, `Amorphous`, `Unknown`
- `ConfigTag` members: `Dendrite`, `Interdendritic`, `Equiaxed`, `Columnar`, `Eutectic`, `Coring`, `Lath`, `Martensite`, `Acicular`, `Lamellar`, `Widmanstatten`, `Matrix`, `Precipitate`, `Intragranular`, `Intergranular`, `Segregation`, `Twin`, `Subgrain`, `Structure`, `Unknown`
- `MeasurementStatistic` members: `mean`, `median`, `lower`, `upper`, `percentile`
- Units registry: `ureg` (a pint `UnitRegistry` instance)
- Pre-defined units: `HV`, `GigaPascal`, `MegaPascal`, `Micrometer`, `Nanometer`, `gram_per_cm3`, `percent`, `dimensionless`, `Celsius`, `Kelvin`, `Atm`
- `ureg` supports accessing any standard pint unit as an attribute (e.g. `ureg.angstrom`, `ureg.joule`, `ureg.meter`).
- You can compose new units with arithmetic: `ureg.megapascal * ureg.meter ** 0.5`, `ureg.gram / ureg.cm ** 3`, `ureg.ampere / ureg.cm ** 2`.
- You can define entirely new units with `ureg.define("HV = 9.807 * megapascal = vickers_hardness")` if a unit is not already available.
- Composition parser: `Composition` (from `pymatgen`)
- Description group: `AlloyDescriptionGroup`
- `MeasurementMethod` members: `XRD`, `DSC`, `TensileTest`, `CompressionTest`, `VickersHardnessTest`, `NanoindentTest`, `ArchimedesMethod`, `OpticalMicroscope`, `SEM`, `TEM`, `STEM`, `EBSD`, `UniversalTestingMachine`, `ResonanceUltrasoundSpectroscopy`, `FractureToughnessTest`, `Balance`, `EDS`, `TEM_EDS`, `WDS`, `EPMA`, `LIBS`, `ED_XRF`, `WD_XRF`, `Spark_OES`, `ICP_OES`, `ICP_MS`, `Unspecified`
- Composition helper functions: `composition_with_weight_additions`, `balance_composition`
- For weight-percent composition dictionaries, prefer `Composition.from_weight_dict(...)` when available in your environment
\end{prompt}

To facilitate canonicalization, we teach the LLM to use the normalization function, which maps a string from the paper to its canonical value.

\begin{prompt}[]
Normalize function:
- `normalize(val, val_in_paper, source=None)` is a documentation wrapper that records when a paper's terminology differs from our standardized value. It returns `val` unchanged.
- Use it whenever the paper uses different terminology than our standardized enum values or process kind strings.
- `val`: the standardized/ground-truth value you want to use.
- `val_in_paper`: the exact term the paper used (as a string).
- `source`: optional string noting where in the paper the term appears.

When to use `normalize`:
- A measurement kind in the paper doesn't exactly match an `AlloyMeasurementKind` member name (e.g. the paper says "Yield Strength" but the measurement was done via compression, so the correct kind is `AlloyMeasurementKind.yield_strength_compression`).
- A process description in the paper is more specific than a `ProcessKind` member (e.g. the paper says "Vacuum Arc Melting" but the correct kind is `ProcessKind.ArcMelting`).

Examples:
```python
# Paper says "Yield Strength" but the test was compressive
Measurement(
  kind=normalize(AlloyMeasurementKind.yield_strength_compression, val_in_paper="Yield Strength"),
  value=1200.0,
  unit=MegaPascal,
)

# Paper says "Fracture Strain" but it is ultimate strain in compression
Measurement(
  kind=normalize(AlloyMeasurementKind.ultimate_strain_compression, val_in_paper="Fracture Strain"),
  value=25.0,
  unit=percent,
)

# Paper says "Vacuum Arc Melting"; the correct ProcessKind is ArcMelting
ProcessEvent(
  kind=normalize(ProcessKind.ArcMelting, val_in_paper="Vacuum Arc Melting"),
  description="Melted under vacuum",
)
```

When NOT to use `normalize`:
- If the paper's terminology already matches the standardized value exactly, just use the value directly (e.g. `kind=AlloyMeasurementKind.vickers_hardness` when the paper says "Vickers hardness").
\end{prompt}

\subsection*{Composition Extraction Task Prompt}\label{appendix:comp_extract_prompt}
Below is the prompt for the composition extraction task outlined in \autoref{section:extracting_compositions}. For the runs where the helper functions were not present, mentions of those functions were omitted.
\begin{prompt}[]
Extract all compositions of alloys/materials that the authors physically synthesized in this paper.

We are only interested in the compositions of alloys/materials that the authors physically made (synthesized) in their lab.
Do NOT include:
- Compositions mentioned in passing from other studies or references
- Compositions from computational/theoretical predictions
- Compositions of raw materials or precursors (e.g. pure elements)
- Compositions of individual phases within a material (e.g. dendrite vs inter-dendrite regions)
Only include the overall composition of each material the authors created.

IMPORTANT: For each material, report only one composition - the one measured by the highest-resolution analytical technique available. Prefer analytically measured compositions over nominal/intended ones. The priority order from best to worst is:
ICP-MS > ICP-OES > WD-XRF > EPMA > WDS > ED-XRF > Spark-OES > EDS > LIBS > nominal/Balance
For example, if a paper reports both a nominal composition and an EDS-measured composition for the same material, use the EDS-measured one. If it reports both EDS and XRF for the same material, use the XRF measurement.

Note: You only have access to the text of the paper. Images, figures, and tables rendered as images are not available, so rely solely on the textual content.

Return exactly one Python fenced code block and nothing else.
The code must set a `result` variable to a `list[Composition]`.
Each entry should be a `pymatgen.core.Composition` object representing one alloy the authors made.

Example:
```python
result = [
  Composition("MgFeNi"),
  Composition("Al0.5CoCrFeNi"),
  Composition.from_weight_dict({"Ni": 60, "Co": 20, "Cr": 20}),
  balance_composition("Ti", {"Al": 6, "V": 4}),
]
```

Available names in runtime:
- `Composition` (from `pymatgen.core`) - use to create composition objects
- `Composition("MgFeNi")` - from formula string (atomic ratio style)
- `Composition.from_weight_dict({"Ni": 60, "Co": 20, "Cr": 20})` - from weight-percent dictionary
- Composition helper functions: `balance_composition`, `composition_with_weight_additions`

Composition helper functions:

1. `balance_composition(main_element, additions)` - for "balance notation" compositions.
 Use when the paper writes compositions like Ti-6Al-4V, meaning the main element (Ti) makes up
 the balance (remainder to 100 wt%) after accounting for the other additions (6 wt% Al, 4 wt% V).
 - `main_element`: string name of the balance element (e.g. `"Ti"`).
 - `additions`: dict mapping element names to their weight percentages (e.g. `{"Al": 6, "V": 4}`).
 - Returns a `Composition` object. Wrap the result in `CompositionMeasurement(...)` before adding to `Material.measurements`.
 - Example: `balance_composition("Ti", {"Al": 6, "V": 4})` - Ti-6Al-4V (Ti is 90 wt% balance).
 - Example: `balance_composition("Fe", {"C": 0.4, "Mn": 1.5})` - Fe balance with 0.4 wt% C and 1.5 wt% Mn.

2. `composition_with_weight_additions(base, additions, addition_wt_frac)` -
 for when the paper says "add X wt% of Y to base alloy".
 - `base`: the original alloy composition before additions (usually atomic-fraction style).
 - `additions`: the additive recipe expressed by weight ratio; use `Composition.from_weight_dict(...)` for this.
 - `addition_wt_frac`: decimal fraction of additive mass relative to base mass (`5 wt% -> 0.05`, `2.5 wt% -> 0.025`).
  Do not pass percent values like `5`; pass decimal fractions like `0.05`.
 - Returns a `Composition` object. Wrap the result in `CompositionMeasurement(...)` before adding to `Material.measurements`.
 - Example: `composition_with_weight_additions(Composition("NbTaTiZr"), Composition.from_weight_dict({"Mo": 50, "W": 50}), 0.05)` - 5 wt% (Mo/W mix) added to NbTaTiZr.

Also available:
- `Composition.from_weight_dict({"Ni": 60, "Co": 20, "Cr": 20})` - create a Composition directly from weight-percent values.

Composition formatting guidance:
- Use plain formulas for atomic-fraction style entries (e.g. `Composition("Nb0.25Ta0.25Ti0.25Zr0.25")`).
- Never append text like `(at.%)` or `(wt.%)` inside composition formula strings.
\end{prompt}

The following prompt is for the string-only composition extraction task described in \autoref{section:extracting_compositions}. It lacks most of the code-specific instructions and only asks the LLM to output a string of compositions that are parsed by Pymatgen. Retries are performed if the extracted composition raises an error when passed into Pymatgen's Composition constructor.

\begin{prompt}[]
Extract all compositions of alloys/materials that the authors physically synthesized in this paper.

We are only interested in the compositions of alloys/materials that the authors physically made (synthesized) in their lab.
Do NOT include:
- Compositions mentioned in passing from other studies or references
- Compositions from computational/theoretical predictions
- Compositions of raw materials or precursors (e.g. pure elements)
- Compositions of individual phases within a material (e.g. dendrite vs inter-dendrite regions)
Only include the overall composition of each material the authors created.

IMPORTANT: For each material, report only one composition - the one measured by the highest-resolution analytical technique available. Prefer analytically measured compositions over nominal/intended ones. The priority order from best to worst is:
ICP-MS > ICP-OES > WD-XRF > EPMA > WDS > ED-XRF > Spark-OES > EDS > LIBS > nominal/Balance
For example, if a paper reports both a nominal composition and an EDS-measured composition for the same material, use the EDS-measured one. If it reports both EDS and XRF for the same material, use the XRF measurement.

Note: You only have access to the text of the paper. Images, figures, and tables rendered as images are not available, so rely solely on the textual content.

Return one composition formula per line inside a fenced code block.
Each formula must be a valid chemical formula parseable by pymatgen's `Composition()` constructor.
Use standard chemical formula notation (e.g. atomic ratio style).

For weight-percent compositions, convert to the equivalent atomic-ratio formula.

Example:
```
Al0.5CoCrFeNi
MgFeNi
Fe2O3
```
\end{prompt}

\subsection*{Single Property Extraction Task Prompt}\label{appendix:single_property_extraction_prompt}
Below is the prompt for the single property extraction task outlined in \autoref{extract_individual_properties}. As seen in the prompt, all properties have a canonical unit, so the model must output the value in the expected unit.
\begin{prompt}[]
Extract all Ultimate tensile strength (UTS) values of alloys/materials that the authors physically synthesized in this paper. Return values in MPa.

We are only interested in measurements of alloys/materials that the authors physically made (synthesized) in their lab.
Do NOT include:
- Values mentioned in passing from other studies or references
- Values from computational/theoretical predictions
- Values of individual phases within a material (e.g. dendrite vs inter-dendrite regions)
Only include the overall material-level values.

Note: You only have access to the text of the paper. Images, figures, and tables rendered as images are not available, so rely solely on the textual content.

Return a JSON array of numeric values. Only include the numeric values, no units or labels.
If there are no values for this property, return an empty array.
If a value exists for some materials, but is missing from others, just include the ones that exist, don't use filler values like -1 or 0.
\end{prompt}

\subsection*{JSON Extraction Format Prompt}\label{appendix:json_extraction_format_prompt}

\autoref{appendix:json_extraction} evaluates whether the output extraction format expressed in JSON would yield different results than the output expressed as code. The resulting prompt for this experiment is quite similar to the prompt for the standard experiment extraction prompt, especially because many subprompts are shared. The main difference is that the output format is JSON, and helper functions are tested using JSON-specific schema definitions.

\begin{prompt}[]
Extract experiments from this paper.

We are only interested in materials the authors physically made in their lab.
Extract ALL properties reported for those materials - mechanical, thermal, physical, microstructural, etc.
Include properties derived from measurements (e.g. strain hardening exponent from a stress-strain curve, fracture toughness from a fracture test, Pugh ductility ratio from measured elastic constants).
Exclude properties that are purely computed from composition or thermodynamic databases (e.g. CALPHAD/ThermoCalc predictions, DFT-computed values, rule-of-mixtures estimates with no experimental validation).

Return a JSON fenced code block (```json ... ```) containing an array of experiment objects.

Use this exact shape:

```json
[
  {
    "raw_materials": {
      "elements": {
        "kind": "Ingot",
        "description": "...",
        "source": "..."
      }
    },
    "synthesis_groups": {
      "creation": [
        {"kind": "ArcMelting", "description": "...", "source": "..."},
        {"kind": "AsCast", "description": "...", "source": "..."}
      ],
      "annealing[Temp]": [
        {
          "kind": "Annealing",
          "temperature": {"value": "[Temp]", "unit": "celsius"},
          "source": "..."
        }
      ]
    },
    "descriptions": [
      {
        "kinds": ["vickers_hardness"],
        "method": "VickersHardnessTest",
        "desc": "The microhardness (HV) of the alloys was measured with a Vickers hardness tester under a load of 500 gf"
      },
      {
        "kinds": ["Grinding"],
        "desc": "performed on a grinding wheel"
      }
    ],
    "output_materials": [
      {
        "process": "elements->creation",
        "name": "materialA",
        "measurements": [
          {"_type": "composition", "composition": "MgFeNi"},
          {
            "_type": "measurement",
            "kind": "vickers_hardness",
            "value": 321.0,
            "unit": "HV",
            "uncertainty": 7.0,
            "source": "..."
          },
          {
            "_type": "measurement",
            "kind": "yield_strength_tension",
            "value": 450.0,
            "unit": "MegaPascal",
            "measurement_statistic": "mean",
            "source": "..."
          },
          {
            "_type": "group_measurements",
            "kind": "grain_size",
            "unit": "micrometer",
            "values": [
              {"statistic": "lower", "value": 5},
              {"statistic": "upper", "value": 50}
            ]
          },
          {
            "_type": "lattice_param",
            "lattice": {"type": "cubic", "a": 3.208},
            "struct": "BCC",
            "source": "..."
          },
          {
            "_type": "configuration",
            "name": "dendrite",
            "tags": ["Dendrite"],
            "measurements": [
              {
                "_type": "composition",
                "composition": {"Mo": 27.6, "Nb": 25.0, "Ta": 31.5, "V": 15.9},
                "method": "EDS",
                "source": "..."
              }
            ]
          },
          {
            "_type": "configuration",
            "name": "FCC matrix",
            "struct": "FCC",
            "tags": ["Matrix"],
            "measurements": [
              {
                "_type": "measurement",
                "kind": "grain_size",
                "value": "~0.71",
                "unit": "micrometer"
              }
            ]
          },
          {
            "_type": "configuration",
            "name": "B2 precipitates",
            "struct": "B2",
            "tags": ["Precipitate", "Intragranular"],
            "within": "FCC matrix",
            "measurements": [
              {
                "_type": "group_measurements",
                "kind": "grain_size",
                "unit": "nanometer",
                "values": [
                  {"statistic": "lower", "value": 50},
                  {"statistic": "upper", "value": 180}
                ]
              }
            ]
          }
        ]
      },
      {
        "process": "materialA->annealing[Temp=10]",
        "name": "materialB",
        "measurements": [
          {"_type": "composition", "composition": "MgFeNi"},
          {
            "_type": "measurement",
            "kind": "vickers_hardness",
            "value": 350.0,
            "unit": "HV",
            "uncertainty": 7.0,
            "source": "..."
          }
        ]
      }
    ]
  }
]
```

How to write each experiment object and fields to populate:
1) "raw_materials" (required): map each initial input name (e.g. "elements" or "powders") to a raw material object.
   - "kind": one of the RawMaterialKind values (usually "Ingot", "Powder", or "Unspecified").
   - Populate "description" and "source" whenever the paper states purity, supplier, or precursor details.
2) "synthesis_groups" (required): an object mapping named synthesis stages to arrays of process event objects.
   - Use reusable stages and process variables when appropriate (e.g. "annealing[Temp]").
   - Each process event MUST include "kind" (a ProcessKind member name). Optionally include "temperature", "duration", "description", "source" when available. If you include temperature or duration, each MUST be a quantity object with BOTH "value" and "unit" (e.g. {"value": 700, "unit": "celsius"}). Omit the field entirely if unknown - never provide only "value" or only "unit".
   - Use "inputs" to declare which raw materials or named materials feed into a specific
     process event. Example: {"kind": "Mixing", "inputs": ["elements", "wc_additions"]}
   - Every raw material declared in "raw_materials" must be referenced somewhere - either as a
     comma-separated input in a material's "process" string or via a process event's "inputs".
   - "ArcMelting" and "InductionMelting" must always be immediately followed by a casting step
     (e.g. "AsCast", "SuctionCasting", "DropCasting", "GravityCasting", "DirectionalSolidification", or "CastingUnspecified").
   - IMPORTANT: Every synthesis group you define MUST be referenced by at least one material's "process" string.
     Do not define groups that are not used.
   - IMPORTANT: If a group name contains a template variable like [Temp] (e.g. "annealing[Temp]"),
     that exact placeholder string "[Temp]" MUST appear as a value in at least one process event field
     within that group. For example: "temperature": {"value": "[Temp]", "unit": "celsius"}.
     Do NOT put the actual number in the group definition - the actual value goes in the material's
     process string (e.g. "process": "elements->annealing[Temp=700]").
   - Only include actual fabrication/processing steps in synthesis_groups.
     Do NOT put measurement methods, characterization techniques, or testing procedures here.
3) "output_materials" (required): array of material objects.
   - "process": use process notation such as "elements->creation" or "base->annealing[Temp=700]->quenching".
   - The first segment (before the first "->") is a comma-separated list of input raw materials
     or named materials. Use commas to combine multiple inputs: "elements,reinforcement->mixing->sintering".
   - Include "name" only if the paper names that material.
   - "measurements" must include at least one composition measurement ({"_type": "composition", "composition": "..."}).
   - Add ALL reported property measurements.
4) Measurements - each item in the "measurements" array must have a "_type" field:
   - "_type": "composition" - for composition. Include "composition" (formula string or element dict) and optionally "method".
   - "_type": "measurement" - for a single measurement. REQUIRED: "kind", "value", "unit" (all three must be present). Optional: "uncertainty", "measurement_method", "temperature", "pressure", "measurement_statistic".
   - "_type": "group_measurements" - for ranges (e.g. "5-50 um"). REQUIRED: "kind", "unit", "values" (array of {"statistic": "lower"/"upper", "value": ...}). All three must be present.
   - "_type": "lattice_param" - for XRD lattice parameters. Include "lattice" ({"type": "cubic", "a": 3.208}), "struct" (e.g. "BCC").
   - "_type": "configuration" - for microstructural features. Include "name", "tags", and nested "measurements".
   - If uncertainty is reported (e.g. "450 +- 20"), set "value": 450.0 and "uncertainty": 20.0.
   - If temperature or pressure is tied to a measurement, set "temperature": {"value": ..., "unit": "celsius"}. Both "value" and "unit" are REQUIRED in any quantity object.
   - Assume room temperature is ~23 C when the paper says "room temperature".
   - Pay attention to how the paper qualifies numeric values:
     - Exact value: "value": 50.0
     - Approximate value: "value": "~50"
     - Greater than: "value": ">50"
     - Less than: "value": "<50"
     - Greater than or equal: "value": ">=50"
     - Less than or equal: "value": "<=50"
     - Much greater than: "value": ">>50"
     - Much less than: "value": "<<50"
   - When the paper explicitly states a value is a mean or average, set "measurement_statistic": "mean".
     Leave it out when the paper does not specify.
5) Lattice parameters (for XRD-determined crystal structure):
   - Use "_type": "lattice_param" with a "lattice" object. Required parameters depend on type:
     - "cubic": {"type": "cubic", "a": ...} (requires "a")
     - "hexagonal": {"type": "hexagonal", "a": ..., "c": ...} (requires "a" and "c")
     - "tetragonal": {"type": "tetragonal", "a": ..., "c": ...} (requires "a" and "c")
     - "orthorhombic": {"type": "orthorhombic", "a": ..., "b": ..., "c": ...} (requires "a", "b", and "c")
   - "struct": the crystal structure (e.g. "BCC", "FCC").
   - "phase_fraction": optionally a quantity object (with both "value" and "unit") for the phase fraction if reported.
   - "name": optional name for the phase (e.g. "BCC phase", "sigma phase").
   - Include "source" noting where in the paper the data appears.
6) Configuration (for microstructural features):
   - Use "_type": "configuration" to describe dendrites, precipitates, phases, lamellae, or regions with distinct microstructure.
   - Do NOT use configuration merely to record where on the bulk material a measurement was taken.
   - "name": identifies the feature (e.g. "dendrite", "FCC matrix", "B2 precipitates").
   - "struct": crystal structure if known (e.g. "BCC", "FCC", "B2").
   - "tags": array of ConfigTag values (e.g. ["Dendrite"], ["Precipitate", "Intragranular"]).
   - "within": reference the "name" of another configuration in the same material for nesting.
     All precipitates MUST have a "within" field referencing the configuration they are contained in.
   - "measurements": array of measurements specific to this feature.
7) "descriptions" (optional): array of description group objects for recording contextual information about
   measurement methods and equipment, or process-related descriptions.
   - Use this for information about HOW measurements were performed (instruments, testing conditions).
   - "kinds": array of AlloyMeasurementKind, PhaseMeasurementKind, ProcessKind, or MeasurementMethod values.
   - "method": optional MeasurementMethod value (string).
   - "desc": free-text description.
   - Example:
     ```json
     {"kinds": ["vickers_hardness"], "method": "VickersHardnessTest",
      "desc": "Microhardness measured with Vickers hardness tester at 500 gf load"}
     ```

Available string values for JSON fields:
- Measurement "_type" values: "composition", "measurement", "group_measurements", "lattice_param", "configuration"
- AlloyMeasurementKind values (for "kind" field): "vickers_hardness", "berkovich_hardness", "pugh_ductility_ratio", "density", "yield_strength_tension", "ultimate_strain_tension", "ultimate_tensile_strength", "fracture_strain_tension", "fracture_strength_tension", "strain_hardening_exponent_tension", "poissons_ratio_tension", "fracture_energy_tension", "true_stress_tension", "yield_strength_compression", "ultimate_strain_compression", "ultimate_compressive_strength", "fracture_strain_compression", "fracture_strength_compression", "strain_hardening_exponent_compression", "poissons_ratio_compression", "fracture_energy_compression", "true_stress_compression", "elastic_limit_compression", "elastic_limit_tension", "youngs_modulus", "fracture_toughness", "work_of_fracture", "crystallite_size", "lattice_strain", "melting_point", "solidus", "liquidus"
- PhaseMeasurementKind values (for Configuration/GlobalLatticeParam measurements): "volume_fraction", "length", "grain_size", "phase_size"
- ProcessKind values: "Mixing", "MechanicalAlloying", "PlanetaryMilling", "GasAtomization", "ArcMelting", "InductionMelting", "CastingUnspecified", "AsCast", "GravityCasting", "DropCasting", "SuctionCasting", "DirectionalSolidification", "SparkPlasmaSintering", "HotPressingSintering", "VacuumFurnace", "Homogenization", "Annealing", "NonIsothermalAnnealing", "IsothermalHolding", "WaterQuenching", "SolutionHeatTreatment", "HotExtrusion", "HotRolling", "ColdRolling", "CrossRolling", "ColdForging", "Press", "FrictionStirProcessing", "ElectricalDischargeMachining", "Cut", "Grinding", "Polishing", "Etching", "AquaRegia", "SandBlasting", "Degreased", "UltrasonicBath", "AirDrying"
- RawMaterialKind values: "Ingot", "Powder", "Plate", "Unspecified", "Other"
- CrysStruct values: "FCC", "BCC", "HCP", "DHCP", "Diamond", "L12", "L10", "B2", "D019", "D03", "Heusler", "Rocksalt", "Zincblende", "C14", "C15", "Perovskite", "Amorphous", "Unknown"
- ConfigTag values: "Dendrite", "Interdendritic", "Equiaxed", "Columnar", "Eutectic", "Coring", "Lath", "Martensite", "Acicular", "Lamellar", "Widmanstatten", "Matrix", "Precipitate", "Intragranular", "Intergranular", "Segregation", "Twin", "Subgrain", "Structure", "Unknown"
- MeasurementStatistic values: "mean", "median", "lower", "upper", "percentile"
- Unit strings: "HV", "GigaPascal", "MegaPascal", "micrometer", "nanometer", "gram_per_cm3", "percent", "dimensionless", "celsius", "kelvin", "atm"
  - You can also use standard unit abbreviations: "GPa", "MPa", "um", "nm", "mm", etc.
- Composition: use a formula string (e.g. "CoCrFeNi") or an element dict (e.g. {"Co": 25, "Cr": 25, "Fe": 25, "Ni": 25})
- Lattice types: "cubic", "hexagonal", "tetragonal", "orthorhombic"
- Description group: object with "kinds", "method", "desc"
- MeasurementMethod values (for "measurement_method" field): "XRD", "DSC", "TensileTest", "CompressionTest", "VickersHardnessTest", "NanoindentTest", "ArchimedesMethod", "OpticalMicroscope", "SEM", "TEM", "STEM", "EBSD", "UniversalTestingMachine", "ResonanceUltrasoundSpectroscopy", "FractureToughnessTest", "Balance", "EDS", "TEM_EDS", "WDS", "EPMA", "LIBS", "ED_XRF", "WD_XRF", "Spark_OES", "ICP_OES", "ICP_MS", "Unspecified"
- Composition helpers: use "_helper" objects (see below) instead of formula strings when needed
- Omit fields that are null/not applicable - do not include them with null values.

Composition helper objects (use these instead of formula strings when needed):

1. Balance composition - for "balance notation" (e.g. Ti-6Al-4V):
   ```json
   {"_helper": "balance_composition", "main_element": "Ti", "additions": {"Al": 6, "V": 4}}
   ```
   Ti is the balance element (90 wt%), Al is 6 wt%, V is 4 wt%.

2. From weight dict - create composition from weight percentages:
   ```json
   {"_helper": "from_weight_dict", "weights": {"Ni": 60, "Co": 20, "Cr": 20}}
   ```

3. Weight additions - add X wt% of a mix to a base alloy:
   ```json
   {"_helper": "weight_additions", "base": "NbTaTiZr", "additions_weights": {"Mo": 50, "W": 50}, "fraction": 0.05}
   ```
   Adds 5 wt% of a 50/50 Mo/W mix to equiatomic NbTaTiZr.
   "fraction" is a decimal: 5 wt% = 0.05, 2.5 wt% = 0.025.

Use these helpers inside the "composition" field of a composition measurement:
```json
{"_type": "composition", "composition": {"_helper": "balance_composition", "main_element": "Fe", "additions": {"Cr": 20}}}
```

Composition formatting guidance:
- Use plain formulas for atomic-fraction style entries (e.g. "Nb0.25Ta0.25Ti0.25Zr0.25").
- Never append text like "(at.%)" or "(wt.%)" inside composition formula strings.
\end{prompt}

\subsection*{Match Paper to Redacted Experiment Task Prompts}\label{appendix:match_paper_to_correct_experiment}

Below are the prompts used to empirically evaluate whether LLMs can match a paper's text to the corresponding output experiments. We compare the performance when the output format contains structured experiments in the form of graph dependencies between samples versus if all samples appear independent from one another. We also ran an experiment where we passed all 19 papers to the LLM and asked it to identify the paper corresponding to a redacted experiment. However, this experiment failed as the text of all 19 papers exceeded the context length of smaller models.

This prompt is used for the task that matches a paper to a set of redacted flat experiments.

\begin{prompt}[]
You are given a scientific paper and 19 redacted experiment summaries (labeled Set 1 through Set 19). Each summary describes the structural fingerprint of the experiments extracted from one of 19 papers. Exactly one summary corresponds to this paper.

Your task: determine which set matches this paper based on structural clues only (counts of raw materials, output materials, measurements, process steps, and their relationships).

Respond with a JSON object: {"reason": "<brief explanation of why you chose this set>", "match": N} where N is the set number (1-19).


--- REDACTED EXPERIMENT SUMMARIES ---

### Set 1
Experiment 1:
  Raw materials: raw_material_1
  Output materials: 5
  raw_material_1 -> (2 synthesis steps) -> material_1:
    1 composition, 0 property measurements, 1 configuration (1 nested measurements), 0 lattice, 0 global lattice
  raw_material_1 -> (2 synthesis steps) -> material_2:
    1 composition, 3 property measurements, 2 configuration (8 nested measurements), 0 lattice, 0 global lattice
  raw_material_1 -> (2 synthesis steps) -> material_3:
    1 composition, 1 property measurements, 2 configuration (7 nested measurements), 0 lattice, 0 global lattice
  raw_material_1 -> (2 synthesis steps) -> material_4:
    1 composition, 1 property measurements, 2 configuration (7 nested measurements), 0 lattice, 0 global lattice
  raw_material_1 -> (2 synthesis steps) -> material_5:
    1 composition, 3 property measurements, 2 configuration (7 nested measurements), 0 lattice, 0 global lattice


...

### Set 19
Experiment 1:
  Raw materials: raw_material_1
  Output materials: 7
  raw_material_1 -> (3 synthesis steps) -> material_1:
    1 composition, 2 property measurements, 4 configuration (1 nested measurements), 0 lattice, 0 global lattice
  raw_material_1 -> (3 synthesis steps) -> material_2:
    1 composition, 2 property measurements, 0 configuration, 0 lattice, 0 global lattice
  raw_material_1 -> (3 synthesis steps) -> material_3:
    1 composition, 1 property measurements, 0 configuration, 0 lattice, 0 global lattice
  raw_material_1 -> (3 synthesis steps) -> material_4:
    1 composition, 1 property measurements, 0 configuration, 0 lattice, 0 global lattice
  raw_material_1 -> (3 synthesis steps) -> material_5:
    1 composition, 1 property measurements, 0 configuration, 0 lattice, 0 global lattice
  raw_material_1 -> (3 synthesis steps) -> material_6:
    1 composition, 1 property measurements, 0 configuration, 0 lattice, 0 global lattice
  raw_material_1 -> (3 synthesis steps) -> material_7:
    1 composition, 1 property measurements, 2 configuration (1 nested measurements), 0 lattice, 0 global lattice


--- PAPER TEXT ---

<paper_text>
\end{prompt}

This prompt is used for the task that matches a paper to a set of redacted experiments organized in a graph manner.

\begin{prompt}[]
You are given a scientific paper and 19 redacted experiment summaries (labeled Set 1 through Set 19). Each summary describes the structural fingerprint of the experiments extracted from one of 19 papers. Exactly one summary corresponds to this paper.

Your task: determine which set matches this paper based on structural clues only (counts of raw materials, output materials, measurements, process steps, and their relationships).

Respond with a JSON object: {"reason": "<brief explanation of why you chose this set>", "match": N} where N is the set number (1-19).


--- REDACTED EXPERIMENT SUMMARIES ---

### Set 1
Experiment 1:
  Raw materials: raw_material_1
  Output materials: 5
  raw_material_1 -> (2 synthesis steps) -> material_1:
    1 composition, 0 property measurements, 1 configuration (1 nested measurements), 0 lattice, 0 global lattice
  raw_material_1 -> (2 synthesis steps) -> material_2:
    1 composition, 3 property measurements, 2 configuration (8 nested measurements), 0 lattice, 0 global lattice
  raw_material_1 -> (2 synthesis steps) -> material_3:
    1 composition, 1 property measurements, 2 configuration (7 nested measurements), 0 lattice, 0 global lattice
  raw_material_1 -> (2 synthesis steps) -> material_4:
    1 composition, 1 property measurements, 2 configuration (7 nested measurements), 0 lattice, 0 global lattice
  raw_material_1 -> (2 synthesis steps) -> material_5:
    1 composition, 3 property measurements, 2 configuration (7 nested measurements), 0 lattice, 0 global lattice


...

### Set 19
Experiment 1:
  Raw materials: raw_material_1
  Output materials: 7
  raw_material_1 -> (3 synthesis steps) -> material_1:
    1 composition, 2 property measurements, 4 configuration (1 nested measurements), 0 lattice, 0 global lattice
  raw_material_1 -> (3 synthesis steps) -> material_2:
    1 composition, 2 property measurements, 0 configuration, 0 lattice, 0 global lattice
  raw_material_1 -> (3 synthesis steps) -> material_3:
    1 composition, 1 property measurements, 0 configuration, 0 lattice, 0 global lattice
  raw_material_1 -> (3 synthesis steps) -> material_4:
    1 composition, 1 property measurements, 0 configuration, 0 lattice, 0 global lattice
  raw_material_1 -> (3 synthesis steps) -> material_5:
    1 composition, 1 property measurements, 0 configuration, 0 lattice, 0 global lattice
  raw_material_1 -> (3 synthesis steps) -> material_6:
    1 composition, 1 property measurements, 0 configuration, 0 lattice, 0 global lattice
  raw_material_1 -> (3 synthesis steps) -> material_7:
    1 composition, 1 property measurements, 2 configuration (1 nested measurements), 0 lattice, 0 global lattice


--- PAPER TEXT ---

<paper_text>
\end{prompt}

\twocolumn

\end{document}